\documentclass[showpacs,aps, twocolumn,nofootinbib]{revtex4-2}
\usepackage{epsfig}
\usepackage{graphicx}
\usepackage{amsmath,amssymb,amsfonts}
\usepackage{array}
\usepackage{url}
\usepackage[usenames,dvipsnames]{color}
\usepackage{hyperref}
\usepackage{multirow}
\usepackage{float}
\usepackage{lineno}
\usepackage{xspace}
\usepackage{cancel}
\usepackage{ulem}
\usepackage{lipsum}

\newcommand{\bef}{\begin{figure}}
\newcommand{\eef}{\end{figure}}
\newcommand{\bc}{\begin{center}}
\newcommand{\ec}{\end{center}}

\newcommand{\be}{\begin{equation}}
\newcommand{\ee}{\end{equation}}
\newcommand{\bea}{\begin{eqnarray}}
\newcommand{\eea}{\end{eqnarray}}

\begin{document}

\title{Impact of strong magnetic field, baryon chemical potential, and medium anisotropy on polarization and spin alignment of hadrons}

\author{Bhagyarathi Sahoo}
\email{Bhagyarathi.Sahoo@cern.ch}
\author{Captain R. Singh}
\email{captainriturajsingh@gmail.com}
\author{Raghunath Sahoo}
\email{Raghunath.Sahoo@cern.ch (Corresponding Author)}
\affiliation{Department of Physics, Indian Institute of Technology Indore, Simrol, Indore 453552, India}

\begin{abstract}
The recent observation of global spin polarization of $\Lambda$ ($\bar{\Lambda}$) hyperons and the spin alignment of $\phi$ and $K^{*0}$ vector mesons create remarkable interest in investigating the particle polarization in the relativistic fluid produced in heavy-ion collisions at GeV/TeV energies. Among other sources of spin polarization phenomena, the Debye mass of a medium plays a crucial role in particle polarization. Any modification brought to the effective mass due to the temperature, strong magnetic field ($eB$), baryonic chemical potential ($\mu_{B}$), medium anisotropy ($\xi$), and vorticity, etc., certainly affects the particle spin polarization. In this work, we explore the global hyperon spin polarization and the spin alignment of vector mesons corresponding to the strong magnetic field, baryonic chemical potential, and medium anisotropy. We find that the degree of spin polarization is flavor-dependent for hyperons. Meanwhile, vector meson spin alignment depends on the hadronization mechanisms of initially polarized quarks and anti-quarks. Medium anisotropy significantly changes the degree of spin polarization compared to the magnetic field and baryon chemical potential.
\end{abstract}
\date{\today}
\maketitle

\section{Introduction}
\label{intro}
So far, the thermalized state of strongly interacting partons, called quark-gluon plasma (QGP), is probed in heavy-ion collisions through a baseline, the $pp$ collisions. It was assumed that QGP existence in $pp$ collisions is next to impossible because it lacks the necessary conditions for the QGP formation. On the contrary, in recent LHC events, ultra-relativistic $pp$ collision experiments have reported behavior similar to heavy-ion collisions, e.g., collective flow, strangeness enhancement, etc.~\cite{ ALICE:2016fzo, CMS:2016fnw}. However, other studies suggest similar phenomena may arise due to the other QCD processes~\cite{OrtizVelasquez:2013ofg, Zhang:2021phk}. These studies raise a question concerning the present QGP signatures and a need for the next generation of probes. In the quest for such a baseline-independent probe, particle spin polarization comes into the picture and can have implications for understanding hot QCD matter. The particle production mechanisms constrain the spin polarization of the hadrons~\cite{Sahoo:2023oid}. However, there are various sources that may lead to the production of polarized hadrons. One such primary source could be the initial quark polarization, which arises due to various phenomena like initial orbital angular momentum (OAM) carried by the colliding nuclei creates a local vorticity and helps to polarize the quarks through spin-orbit coupling~\cite{Liang:2004ph, Liang:2004xn, Liang:2007ma, Gao:2007bc}. Moreover, this vorticity induces the flow of axial current along the axis of rotation known as the chiral vortical effect (CVE), and leads to local polarization of quarks in heavy-ion collisions~\cite{Gao:2012ix, Teryaev:2017wlm}. On the other hand, the magnetic field produced due to the charged spectators interacting with the electric charge, and the spin of the quarks may also lead to quark polarization. Additionally, the local correlations or fluctuations of strong vector meson force field could also lead to quark polarization~\cite{Sheng:2019kmk, Sheng:2022wsy}. Furthermore, the anisotropic expansion of the thermalized QCD medium may induce a local polarization of the final state hadrons~\cite{Becattini:2017gcx, STAR:2019erd}. Therefore, determining the contribution of each possible source of spin polarization in ultra-relativistic collisions is a complex phenomena that requires comprehensive theoretical studies quantifying the impact of all these mechanisms separately. Following the motivation, the present work aims to quantify the role of Debye mass in particle spin polarization in heavy-ion collisions with certain medium conditions. \\

In 2005, Liang and Wang predicted that in non-central heavy-ion collisions, the OAM of the partonic system 
polarizes the quarks and anti-quarks through spin-orbit coupling.
The initial partons are proposed to generate a longitudinal fluid shear distribution representing the local relative OAM in the same direction as global OAM, inducing quark polarization. This quark polarization manifests the polarization of the hadrons (with finite spin) along the direction of OAM during the process of hadronization~\cite{Liang:2004ph, Liang:2004xn, Liang:2007ma, Gao:2007bc}. Apart from global polarization of  hadrons, such global quark polarization has many observable consequences, such as left-right asymmetry in hadron 
spectra and global transverse polarization of thermal photons, dileptons, etc.~\cite{FNAL-E704:1991ovg, 
Ipp:2007ng}. They have studied the global polarization of hyperons~\cite{Liang:2004ph} and spin alignment of vector 
mesons~\cite{Liang:2004xn} in different hadronization scenarios. Following, in 2013, Becattini et al. predicted the global spin 
polarization of $\Lambda$ hyperons due to the OAM-manifested thermal vorticity~\cite{Becattini:2013fla}. Various theoretical 
predictions of global $\Lambda$ hyperons polarization by different hydrodynamic and transport models are well agreed with the 
experimental results available at the Relativistic Heavy Ion Collider (RHIC) energies~\cite{STAR:2007ccu, STAR:2017ckg, STAR:2018gyt, STAR:2021beb, STAR:2023nvo, Karpenko:2016jyx, Jiang:2016woz, Xie:2017upb, Fang:2016vpj, Alzhrani:2022dpi,  
Vitiuk:2019rfv, Sun:2017xhx, Ivanov:2019ern, Ivanov:2022ble, Pradhan:2023rvf,Wei:2018zfb,Becattini:2024uha,Guo:2021udq}. These hydrodynamic and transport models considered the thermal equilibrated distribution of spin degrees of freedom for polarization studies~\cite{Csernai:2013bqa, Csernai:2014ywa, Pang:2016igs, Li:2017slc, Fu:2020oxj, Karpenko:2017lyj, Becattini:2020ngo, Xie:2020}. However, recent observations by the ALICE collaboration reported almost zero spin polarizations for $\Lambda$ hyperons within uncertainties at the Large Hadron Collider (LHC) energies~\cite{ ALICE:2019onw}.\\

In addition to $\Lambda$ hyperons, the averaged global spin polarization measurement of $\Omega$ and $\Xi$ hyperons are obtained using 
the polarization transfer method in STAR  collaboration~\cite{STAR:2020xbm, Alpatov:2020iev}. Similarly, the spin polarization of quarkonium states is obtained by 
measuring the angular distribution of dileptons~\cite{ALICE:2023jad, ALICE:2020iev, STAR:2013iae, STAR:2020igu, Sahoo:2023oid}. 
Furthermore, the global spin alignment of  $\phi$, $K^{*0}$, and $K_{S}^{0}$ mesons is observed at the LHC and RHIC from the angular 
distribution of decay daughters in the mother particles's rest frame~\cite{STAR:2022fan, STAR:2008lcm, Singha:2020qns, 
ALICE:2019aid, Singh:2018uad, Mohanty:2021vbt}. Recent spin alignment measurements for $\phi$ vector mesons by the STAR 
collaboration initiate a spark in the theoretical community to investigate the source of large positive spin alignment for $\phi$ 
vector meson, while $K^{*0}$ shows no such spin alignment within uncertainties~\cite{STAR:2022fan}. In some theoretical studies, it is claimed that in addition to vorticity and electromagnetic fields, a strong vector meson force field is 
responsible for hyperon spin polarization~\cite{Csernai:2018yok} as well as spin alignment of vector mesons~\cite{Sheng:2019kmk, 
Sheng:2022wsy}. Recently, Sheng et al. ~\cite{Sheng:2022wsy} explain the large spin alignment of $\phi$ vector meson in a quark 
polarization mechanism by a strong force field, in which quarks interact with the dense medium through a strong force and become 
polarized. Apart from these sources, there could be various other possible sources for vector meson spin alignment, which require further investigation.\\

Moreover, recent observations of longitudinal spin polarization of  $\Lambda$ hyperons as a function of the azimuthal angle at the LHC and RHIC energies disagree with the theoretical predictions~\cite{STAR:2019erd}, which opens a new door for further exploration in this direction.
However, various studies have been developed to solve this puzzle through different aspects~\cite{Becattini:2021iol, Wu:2019eyi}. Similar to hyperons, the local spin polarization of quarks and anti-quarks due to the anisotropic expansion of the medium contributes to the spin 
alignment of vector meson~\cite{Xia:2020tyd}. In Ref.~\cite{Xia:2020tyd}, the authors have argued that the finite spin alignment is not solely coming from global spin alignment but also has the contribution of quarks and/or anti-quarks, which are locally polarized. The contribution of local quark polarization can be obtained via the off-diagonal elements of the spin density matrix.\\

The experimental measurement of the hyperons and vector mesons spin polarization follows different methodologies. However, both depend on 
the angular distribution of decay daughters in the rest frame of the parent particles. The hyperons undergo parity-violating weak decays, 
while the vector mesons mainly decay through the parity-conserving strong decay. The $\Lambda$ hyperon spin polarization is estimated by 
the projection of the momentum of the daughter proton along the OAM (or vorticity axis) direction. On the other hand, the spin alignment 
of vector meson has been studied by measuring the diagonal elements of 3 $\times$ 3 hermitian spin density matrix ($\rho_{m,m^{'}}$), where $ m $ and $m^{'}$ label the spin component along the quantization axis. The diagonal elements  $\rho_{11}$, $\rho_{00}$, and $\rho_{-1-1}$ define the probabilities of the spin component of vector mesons along the quantization axis. Out of these three diagonal elements, $\rho_{00}$ is independent and can be measured in experiments in two-body decays to pseudoscalar mesons. In the absence of spin alignment, all the three spin states of vector mesons 1, 0, and -1 are equally probable, and thus $\rho_{00}$ = 1/3. The deviation of $\rho_{00}$ from 1/3 (i.e. $\rho_{00} - 1/3$) determines the degree of alignment of vector meson.\\

Ultra-peripheral heavy-ion collisions have several consequences; one such example is that these collisions produce a strong magnetic field ( $eB \sim m_{\pi}^{2} \sim 10^{18} $ G). The strength of the magnetic field may reach up to the order of 0.1$m_{\pi}^{2}$, $m_{\pi}^{2}$, and $15 m_{\pi}^{2}$ for SPS, RHIC, and LHC energies, respectively~\cite{Skokov:2009qp}. This magnetic field decays with time and can, in principle, affect the QGP medium evolution, thermodynamic, and transport properties, and spin polarization of various hadrons~\cite{Skokov:2009qp, Sahoo:2023xnu, Sahoo:2023vkw, Sheng:2020ghv, Yang:2017sdk, Becattini:2016gvu, Wu:2023dfz, Han:2017hdi, Guo:2019joy, Xu:2022hql, Ayala:2020soy, Guo:2019mgh, Gao:2012ix}. The effect of magnetic field on the spin polarization of hadrons is discussed in Ref.~\cite{Yang:2017sdk, Sheng:2020ghv, 
Becattini:2016gvu, Wu:2023dfz, Han:2017hdi, Xu:2022hql, Ayala:2020soy, Guo:2019mgh, Gao:2012ix,Guo:2019joy}. Recent observation of global 
spin polarization of $\Lambda$ 
hyperons indicate that hyperon spin polarization decreases with the increasing center of mass energy at mid-rapidity~\cite{STAR:2007ccu, STAR:2017ckg, 
STAR:2018gyt}. So, at lower energies, the probability of particles getting polarized is large, which could be a consequence of 
baryon stopping. Such high baryon density may affect the evolution process and particle polarization. Further, due to the 
rapid longitudinal expansion of the fireball created in heavy-ion collisions, a high degree of momentum-space anisotropy is produced 
in the QGP medium in its local rest frame~\cite{Dong:2021gnb}. The momentum-space anisotropy has many novel consequences; for 
example, it is essential in modeling heavy-quark dynamics, making the inter-quark potentials spatially anisotropic. Anisotropy present  
in the QGP medium may affect the particle yield, flow coefficients, and particle polarization.\\

In this work, we investigate the effect of the strong magnetic field ($eB$), baryon chemical potential ($\mu_{B}$), and momentum-space anisotropy 
parameter ($\xi$) on the spin polarization of hyperons and spin alignment of vector mesons through Debye screening mass. The separate and combined effects of $eB$, $\mu_{B}$ and 
$\xi$ on the spin polarization and spin alignment of hadrons is explored in the present study. Please note that in the rest of the paper, we shall refer to Debye screening mass as Debye mass. Debye mass is the property of the gauge bosons, calculated from the pole of the dressed gauge boson propagator. In QCD, the temporal component of the gluon self-energy tensor at the static limit gives the Debye mass. In the deconfined medium, it has contributions both from the gluons and quarks degrees of freedom. In the presence of a magnetic field, the magnetic field-dependent contribution to Debye mass arises from the quark loop only since gluons do not interact with the magnetic field. Similarly, the gluon self-energy tensor gets modified in the presence of baryon chemical potential and hence the Debye mass. Furthermore, for an anisotropic QGP case, the anisotropic Debye mass can be factorized into the isotropic Debye mass and anisotropic contribution in a parametrized way~\cite{Dong:2021gnb, Romatschke:2003ms, Romatschke:2004jh, Strickland:2011aa, Dong:2021gnb, Dong:2022mbo, Strickland:lectures}. Here, we have considered two different hadronization 
mechanisms for both hyperons and vector mesons, i.e., the recombination and fragmentation processes. A detailed discussion about hadron polarization for different hadronization scenarios and Debye mass with $eB$, $\mu_{B}$, and $\xi$ is given in Sec.~\ref{formulation}. It is important to mention that the magnetic field may generate 
an additional anisotropy in the medium, which is not explicitly considered here. Further, in section~\ref{res},  we have shown and discussed the results obtained for polarization of hadron under various scenarios. Finally, a brief summary of the work is provided in section~\ref{sum}.

\section{Formulation}

\label{formulation}
The global quark polarization is obtained through parton elastic scattering within the framework of an effective static potential model. This model uses effective mass or Debye mass to incorporate the parton scattering, which induces the quark polarization~\cite{Liang:2004ph, Liang:2004xn, Liang:2007ma, Gao:2007bc}, given as,

\begin{equation}
P_{q} = - \frac{\pi}{2} \frac{m_{D}\bf{p}}{E(E+m_{q})}
\label{Polquark}
\end{equation}

where, $E = \sqrt{{\bf{p}^{2}}+m_{q}^{2}}$, $\bf{p}$ and $m_{q}$ are the energy, momentum, and mass of quark in the 
center of mass frame of the parton scattering, $m_{D}$ is the Debye mass.\\

This global quark polarization leads to the global hyperons spin polarization at the hadronization stage. The spin polarization of hyperons under recombination and fragmentation processes is explicitly reported in the literature~\cite{Yang:2017sdk, Liang:2004ph}. The global spin polarization due to exclusive recombination process for $\Lambda$, $\Sigma^{\pm}$, $\Xi^{-}$, $\Omega^{-}$, and $\Delta^{++}$  hyperons are given as~\cite{Liang:2004ph, Yang:2017sdk}; 

\begin{equation}
P^{rec}_{\Lambda} = P_{s}, \hspace{0.5 cm}       P^{rec}_{\Xi^{-}} = \frac{4P_{s}-P_{d}-3P_{d}P_{s}^{2}}{3-4P_{d}P_{s}+P_{s}^{2}},   
\label{PolLambdarec}
\end{equation}

\begin{eqnarray}
P^{rec}_{\Sigma^{+}} = \frac{4P_{u} -P_{s}-3P_{s}P_{u}}{3- 4P_{u}P_{s}+P_{u}^2}, \nonumber \\  P^{rec}_{\Sigma^{-}} = \frac{4P_{d} -P_{s}-3P_{s}P_{d}}{3- 4P_{d}P_{s}+P_{d}^2} 
\label{Polsigmarec}
\end{eqnarray}

\begin{equation}
P^{rec}_{\Delta^{++}} = \frac{P_{u}(5+P_{u}^{2})}{3(1+P_{u}^{2})}
\hspace{0.5 cm}   P^{rec}_{\Omega^{-}} = \frac{P_{s}(5+P_{s}^{2})}{3(1+P_{s}^{2})}  
\label{Poldeltarec}
\end{equation}

The hyperon spin polarization from inclusive recombination is difficult to estimate. However, at the extreme limit of inclusive recombination (i.e., fragmentation), the hyperon spin polarization due to polarized quarks is estimated in Ref.~\cite{Liang:2004ph}. The spin polarization due to fragmentation process for $\Lambda$, $\Sigma^{\pm}$, $\Xi^{-}$, $\Omega^{-}$, and $\Delta^{++}$ hyperons  are given as;
\begin{equation}
P^{frag}_{\Lambda} = \frac{n_{s}P_{s}}{n_s +2 f_s}, \hspace{0.5 cm} P^{frag}_{\Xi^{-}} = \frac{4n_sP_{s}-f_sP_{d}}{3(2n_s +f_s)} 
\label{Pollambdafrag}
\end{equation}
\begin{equation}
P^{frag}_{\Sigma^{+}} = \frac{4f_sP_{u}-n_sP_{s}}{3(n_s +2 f_s)}  , \hspace{0.15 cm}    P^{frag}_{\Sigma^{-}} = \frac{4f_sP_{d}-n_sP_{s}}{3(n_s +2 f_s)}    
\label{Polcascadefrag}
\end{equation}

\begin{equation}
P^{frag}_{\Omega^{-}} = \frac{P_{s}}{3}, \hspace{0.15 cm}P^{frag}_{\Delta^{++}} = \frac{P_{u}}{3}  
\label{PolOmegafrag}
\end{equation}

Unlike spin polarization for hyperons, the ``twin effect" called the spin alignment of vector meson is described by the $00^{\rm th}$ element of the spin density matrix, i.e., $\rho_{00}$. Such global spin polarization of quarks and anti-quarks also describes the spin alignment of vector mesons for the recombination and fragmentation processes~\cite{Liang:2004xn}. The $00^{\rm th}$ element of spin density matrix for $\rho $, $\phi$ and $K^{*}$ mesons for recombination and fragmentation processes are given as,

\begin{eqnarray}
\rho^{\rho(rec)}_{00} = \frac{1 - P_{q}^2}{3+P_{q}^{2}}, \hspace{0.1 cm}\rho^{\phi(rec)}_{00} = \frac{1 - P_{s}^2}{3+P_{s}^{2}}, \nonumber \\
\rho^{K^{*(rec)}}_{00} = \frac{1 - P_q P_s}{3+P_q P_s}
\label{spinrhorec}
\end{eqnarray}

\begin{equation}
\rho^{\rho(frag)}_{00} = \frac{1 + \beta P_{q}^2}{3- \beta P_{q}^{2}}, \hspace{0.2 cm} \rho^{\phi(frag)}_{00} = \frac{1 + \beta P_{s}^2}{3- \beta P_{s}^{2}}
\label{spinrhofrag}
\end{equation}

\begin{equation}
\rho^{K^{*(frag)}}_{00} = \frac{f_s}{n_s + f_s}  \frac{1 + \beta P_{q}^2}{3- \beta P_{q}^{2}} +  \frac{n_s}{n_s + f_s}  \frac{1 + \beta P_{s}^2}{3- \beta P_{s}^{2}}
\label{spinkstarfrag}
\end{equation}

where $\beta \simeq 0.5$. $n_s$ and  $f_s$ are the strange quark abundances relative to up and down quarks in QGP and quark fragmentation, respectively. \\

In the presence of the magnetic field, the global quark polarization defined in Eq.~\ref{Polquark} modified into, 

\begin{equation}
P_{q} = - \frac{\pi}{2} \sum_{l} \frac{m_{D}\bf{p}}{E_z(E_z + m_{q})}
\label{PolquarkeB}
\end{equation}
and,
\begin{equation}
E_{z} = \sqrt{p_z^{2} + m_{q}^2 + 2l| Q_{i} eB |} \nonumber
\end{equation} 

where $p_z$ is the momentum component along the z direction, $l$ = 0, 1, 2 ... denotes the order of Landau levels, and $Q_{i}e$ is the fractional charge of quark. The detailed derivation of Eq.~\ref{PolquarkeB} is calculated in Appendix~\ref{PqeB}. \\

Further, we have briefly outlined the formalism to explore the effect of magnetic field, baryon chemical potential, and anisotropic parameters on the polarization of hyperons and mesons.

\subsection{Debye Mass under Strong Magnetic Field}

The magnetic field produced in the peripheral heavy-ion collisions is believed to decay quickly. So, it is expected to only affect the partonic phase during the early stage of the evolution process. Magnetic fields tend to align charged particles in a specific direction, resulting in the polarization of the quarks. Given that quarks possess a fractional electric charge, a magnetic field could potentially modify the orientation of quarks within the QGP phase. Consequently, in the presence of a strong magnetic field, the total Debye mass for partons in the static limit has been calculated through the temporal component of one loop vacuum self-energy diagram and has the following form~\cite{Karmakar:2018aig, Zhang:2020efz, Bandyopadhyay:2017cle, Singh:2017nfa};

\begin{equation}
m_{D}^{2}(T, eB) = 4\pi\alpha_{s}(\Lambda^{2}, |eB|)\bigg[T^{2}\frac{N_c}{3} + \sum_{f} \frac{|q_f eB|}{4\pi^{2}}\bigg]
\label{mDeB}
\end{equation}
here, $\alpha_{s}(\Lambda^{2}, |eB|)$  is the running coupling constant, which depends on the magnetic field and temperature. It can be written as~\cite{Bandyopadhyay:2017cle},

\begin{equation}
\alpha_{s}(\Lambda^{2}, |eB|) = \frac{g_{s}^{2}}{4\pi} = \frac{\alpha_{s}(\Lambda^{2})}{1+b_{1}\alpha_{s}(\Lambda^{2})\ln\left(\frac{\Lambda^{2}}{\Lambda^{2} + |eB|}\right)}
\label{alphaeB}
\end{equation}
with
\begin{equation}
\alpha_{s}(\Lambda^{2})  = \frac{1}{b_1\ln\left(\frac{\Lambda^{2}}{\Lambda_{\overline{MS}}^2}\right)} 
\label{alpha}
\end{equation}

where $b_{1}=\frac{11N_c - 2N_f}{12\pi}$, $\Lambda_{\overline{MS}}$ = 0.176 GeV for  $N_f = 3$ and renormalization scale $\Lambda$  = $2 \pi T$~\cite{Haque:2014rua}. \\

In the absence of a magnetic field, the squared of Debye mass is obtained  from the temporal component of the gauge propagator at the static limit and has the following form~\cite{Karmakar:2018aig}
\begin{equation}
m_{D}^2(T) = 4\pi\alpha_{s}(\Lambda^2)T^{2}\left(\frac{N_c}{3} + \frac{N_f}{6} \right)
\label{mD}
\end{equation}

\vspace{1mm}
\subsection{Debye Mass with Baryon Chemical Potential}

Finite $\mu_{B}$ dictates the dominance of the number of quarks over anti-quarks, such an imbalance affects the final production yield of hadrons. In nature, any order parameter or disparity ultimately leads to the polarization of the system. Likewise, a non-zero baryon chemical potential can be the source of a polarized QGP medium. Any change brought to the QGP affects its effective mass, as a consequence, $\mu_{B}$ is incorporated in the Debye mass. In a hot and high baryon density medium (T, $\mu_{B} >> m_{q}$, with $m_q$ as the light quark mass), the leading order HTL/HDL calculation predicts  resultant Debye  mass which has the following form~\cite{Zhao:2022cvl, Lafferty:2019jpr, Huang:2021ysc, Bellac};

\begin{equation}
m_{D}^{2}(T,\mu_{B}) =  4\pi\alpha_{s}(\Lambda^2)\bigg[T^{2}\left(\frac{N_c}{3} + \frac{N_f}{6} \right)   +  \sum_{f} \frac{\mu_{B}^2}{18 \pi^2} \bigg]
\label{mDmuB}
\end{equation}
Here, the strong coupling constant $\alpha_{s}(\Lambda^2)$ takes the form given in Eq.~\ref{alpha}, and in the presence of baryon chemical potential, the renormalization scale is modified into 
\begin{equation}
\Lambda = 2\pi \sqrt{ T^{2}+ \frac{\mu_{B}^2}{\pi^{2}}}
\label{lambda}
\end{equation}

The Debye mass at zero baryon chemical potential can be obtained by putting $\mu_{B}$ = 0 in Eq.~\ref{mDmuB}. 

\subsection{Effect of  Anisotropy on Debye Mass}
Momentum space anisotropy originated due to the collision geometry causing differential expansions of the fireball along the longitudinal and transverse directions. As a consequence, it also alters the effective mass of the medium. The effect of momentum-space anisotropy of the medium on Debye mass is given as~\cite{Romatschke:2003ms, Romatschke:2004jh, Strickland:2011aa, Dong:2021gnb, Dong:2022mbo, Strickland:lectures};

\begin{equation}
m_{D}(T,\xi) =  m_{D}(T) \bigg[  f_{1}(\xi) \cos^2\theta + f_{2}(\xi) \bigg]^{-\frac{1}{4}}
\label{mDzeta}
\end{equation}
and, 
\begin{equation}
m_{D}(T) = \sqrt{4\pi\alpha_{s}(\Lambda^2)T^{2}\left(\frac{N_c}{3} + \frac{N_f}{6} \right)}
\end{equation}
The parameter $\xi$ measures the degree of momentum-space anisotropy, defined as;
\begin{equation}
\xi = \frac{1}{2} \frac{<{\bf{k}_{\perp}^2}>}{<k^2_{z}>} -1
\label{zeta}
\end{equation}
where $k_{z}\equiv\bf{k} \cdot \bf{n}, \bf{k}_{\perp}\equiv \bf{k} - \bf{n}(\bf{k} \cdot \bf{n}$) correspond to the particle momenta along and perpendicular to the direction of anisotropy, and $\theta$ is the angle between particle momentum ($\bf{k}$) with respect to direction of anisotropy ($\bf{n}$). Here, we assume a classical picture in which the direction of anisotropy is along the direction of the particle. The $f_{1}(\xi)$ and $f_{2}(\xi)$ are parameterized\footnote{Here, it is worth mentioning that the parametrization of the anisotropic Debye mass for arbitrary $\xi$ is obtained by constructing a model for the real part of short-range heavy quark-antiquark potential~\cite{Strickland:2011aa}.} in such a way that Eq.~\ref{mDzeta} remains always true under small and large values of $\xi$,

\begin{widetext}
\begin{equation}
f_{1}(\xi) = \frac{9 \xi (1+\xi)^{\frac{3}{2}}}{2\sqrt{3+\xi}(3+\xi^2)} \frac{\pi^2 (\sqrt{2}-(1+\xi)^{\frac{1}{8}} )+ 16(\sqrt{3+\xi}-\sqrt{2})}{(\sqrt{6}-\sqrt{3})\pi^2 - 16(\sqrt{6}-3)}
\label{f1zeta}
\end{equation}

\begin{equation}
f_{2}(\xi) = \xi \bigg(\frac{16}{\pi^2}- \frac{\sqrt{2}(\frac{16}{\pi^2}-1)+(1+\xi)^{\frac{1}{8}}}{\sqrt{3+\xi}} \bigg)\bigg( 1 -  \frac{(1+\xi)^{\frac{3}{2}} }{1 + \frac{\xi^2}{3}}\bigg)+   f_{1}(\xi) +1
\label{f2zeta}
\end{equation}
\end{widetext}

\subsection{Net Debye Mass}
In previous sections, we have discussed the Debye mass for individual cases. Now combining all together ($eB$, $\mu_{B}$, and $\xi$), we have obtained the resultant Debye mass~\cite{Dong:2021gnb};

\begin{equation}
m_{D}(T, \mu_{B}, eB, \xi ) =  m_{D}(T, \mu_{B}, eB) \bigg[  f_{1}(\xi) \cos^2\theta + f_{2}(\xi) \bigg]^{-\frac{1}{4}}
\label{eq14}
\end{equation}
and $m_{D}(T, \mu_{B}, eB)$ is the Debye mass in the presence of both magnetic field and baryon chemical potential calculated from the gluon self-energy tensor in Ref.~\cite{Khan:2021syq}. It has the following form;
\begin{widetext}
\begin{equation}
m_{D}(T, \mu_{B}, eB) = \sqrt{ 4\pi\alpha_{s}(\Lambda^{2}, |eB|)\bigg[T^{2}\frac{N_c}{3}   +  \sum_{f} \frac{|q_f eB|}{4\pi^{2}T} \int_{0}^{\infty} dp_z \{ n^{+}(E_{z})(1- n^{+}(E_{z}))  + n^{-}(E_{z})(1- n^{-}(E_{z})) \}\bigg]}   
\label{eq15}
\end{equation}
\end{widetext} 
with
\begin{equation}
n^{\pm}(E_{z}) = \left[\exp{\frac{( E_{z} \mp \mu_B) }{T}} + 1\right]^{-1}
\end{equation}

The $ n^{+}(E_{z}) $ and $ n^{-}(E_{z}) $ are the phase-space distribution functions of quarks and anti-quarks, respectively. Furthermore, it is important to mention that the Debye mass at the leading order has the same form in the absence of magnetic field, $\mu_{B} \rightarrow 0$, and $\xi \rightarrow 0$ limit. \\


\section{Results and Discussion}
\label{res} 

\begin{figure*}
\includegraphics[scale=0.29]{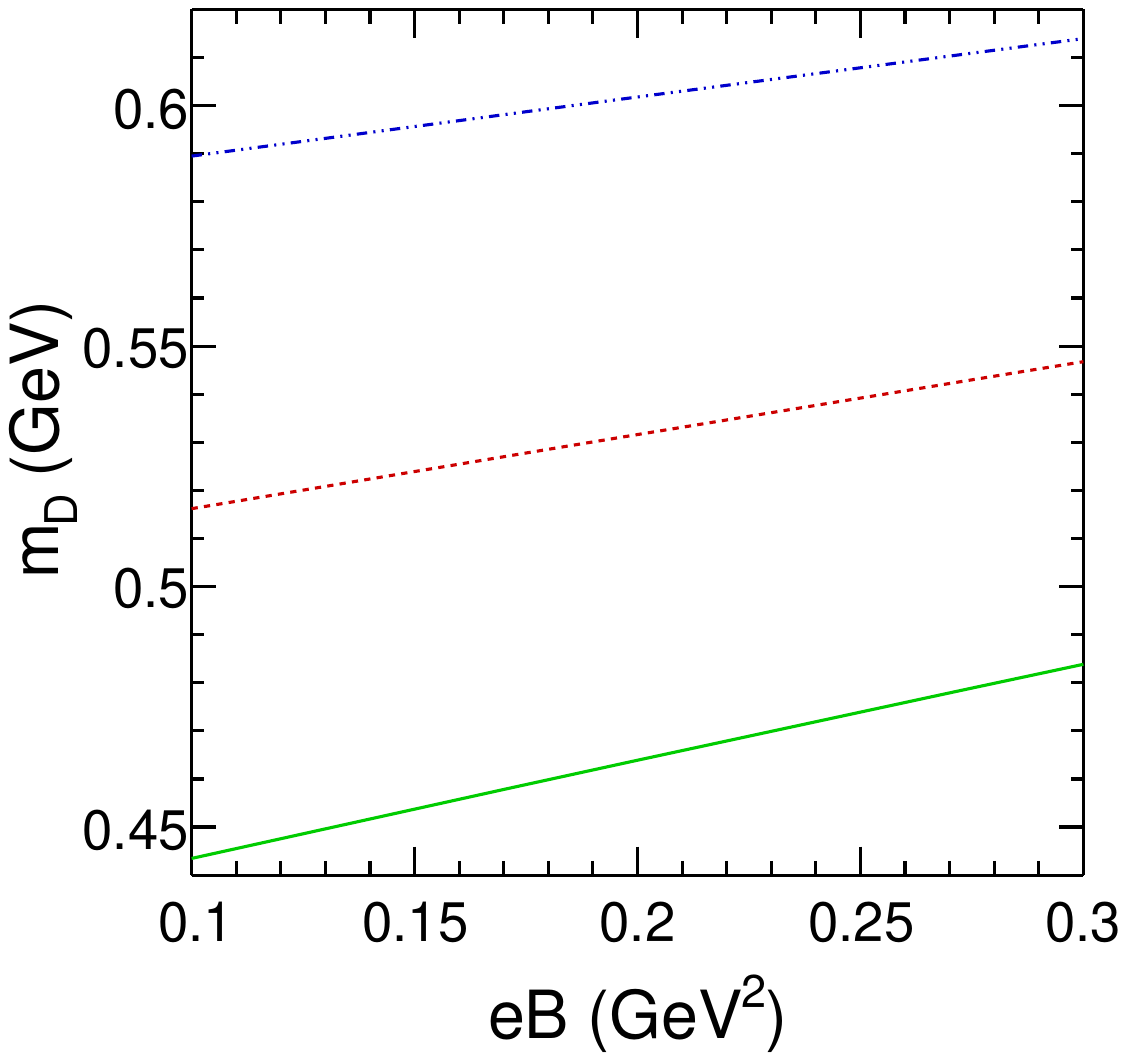}
\includegraphics[scale=0.29]{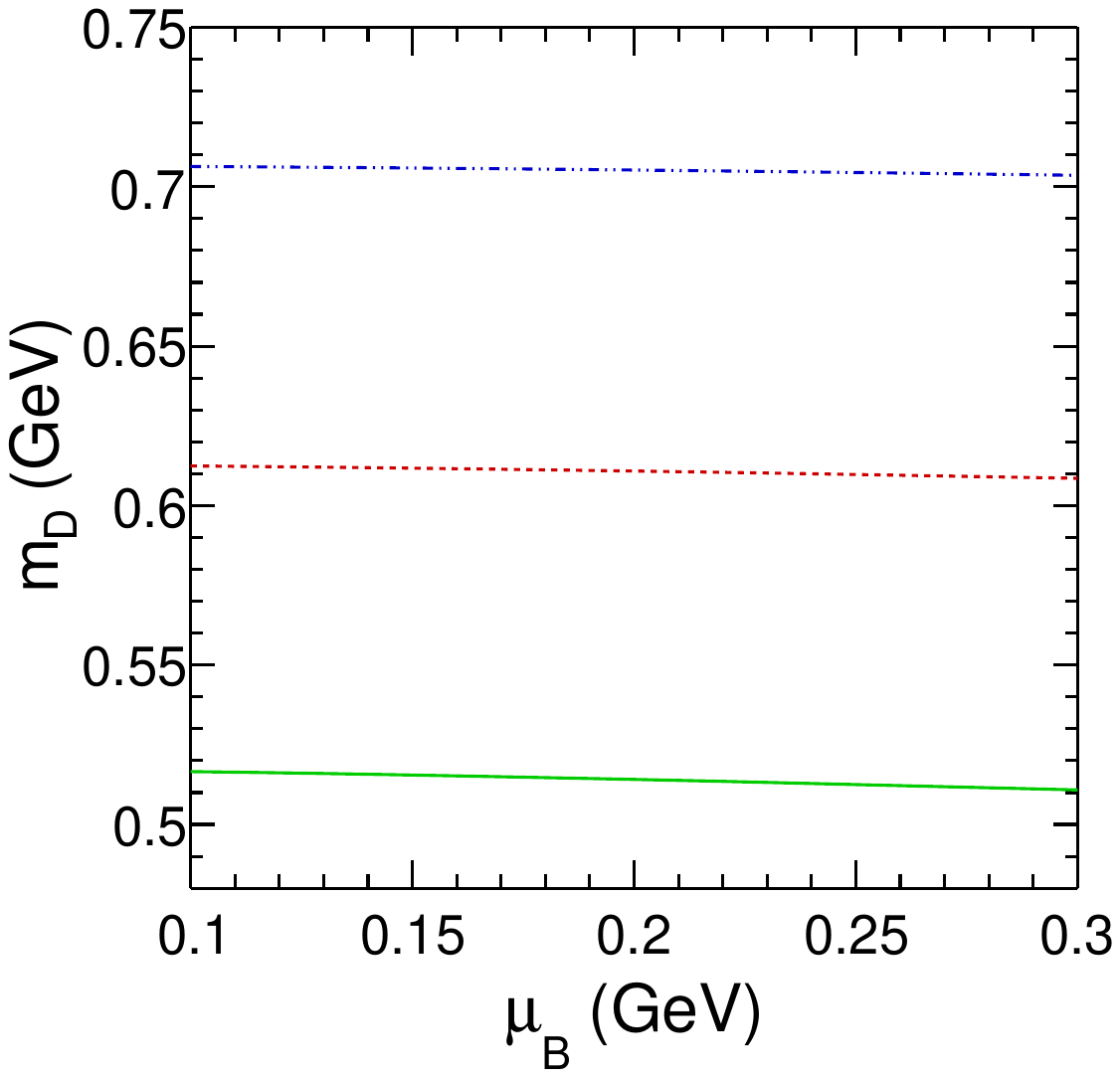}
\includegraphics[scale=0.29]{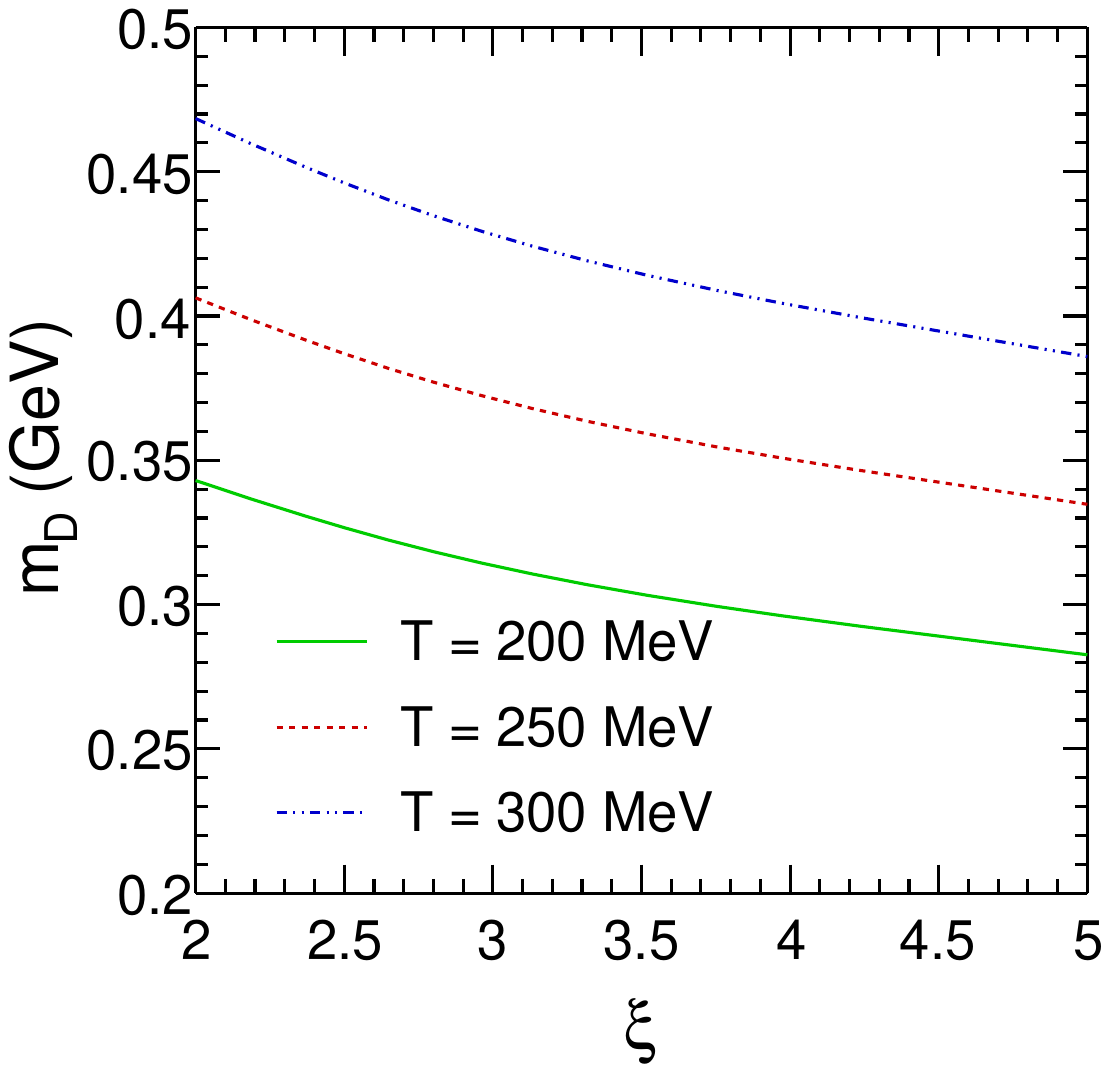} 
\caption{(Color online) The variation of Debye mass ($m_{D}$) as a function of magnetic field ($eB$) (left panel), baryon chemical potential ($\mu_{B}$) (middle panel) and anisotropic parameter ($\xi$) (right panel) for three different values of temperatures, i.e. T = 200 MeV, T=250 MeV, and T = 300 MeV.}
\label{fig:1}
\end{figure*}

The impact of the magnetic field ($eB$), baryon chemical potential ($\mu_{B}$), and anisotropic parameter ($\xi$) on the spin polarization of hyperons and spin alignment of vector mesons are exhibited in this section. Quark-antiquark polarization leads 
to global hadron polarization, and based on that, we have predicted the global spin polarization of various hadrons. Further, the spin polarization of hyperons and the spin alignment 
of vector mesons are obtained for two different hadronization processes: (i) recombination and (ii) fragmentation. Firstly, 
the hyperon spin polarization prediction is given and discussed. Subsequently, the spin alignment of vector mesons is presented. But before 
that, the change in the Debye mass due to $eB$, $\mu_{B}$, and $\xi$ is depicted in Fig.~\ref{fig:1} to ensure the validity of our results. Figure~\ref{fig:1} shows the variation of effective Debye mass with the magnetic field (left panel), baryon chemical potential (middle panel), and anisotropic parameter (right panel) for three different values of temperature, 
i.e., T = 200 MeV, T = 250 MeV, and T = 300 MeV.\\

\begin{figure*}[!htp]
\bc   
\includegraphics[scale=0.29]{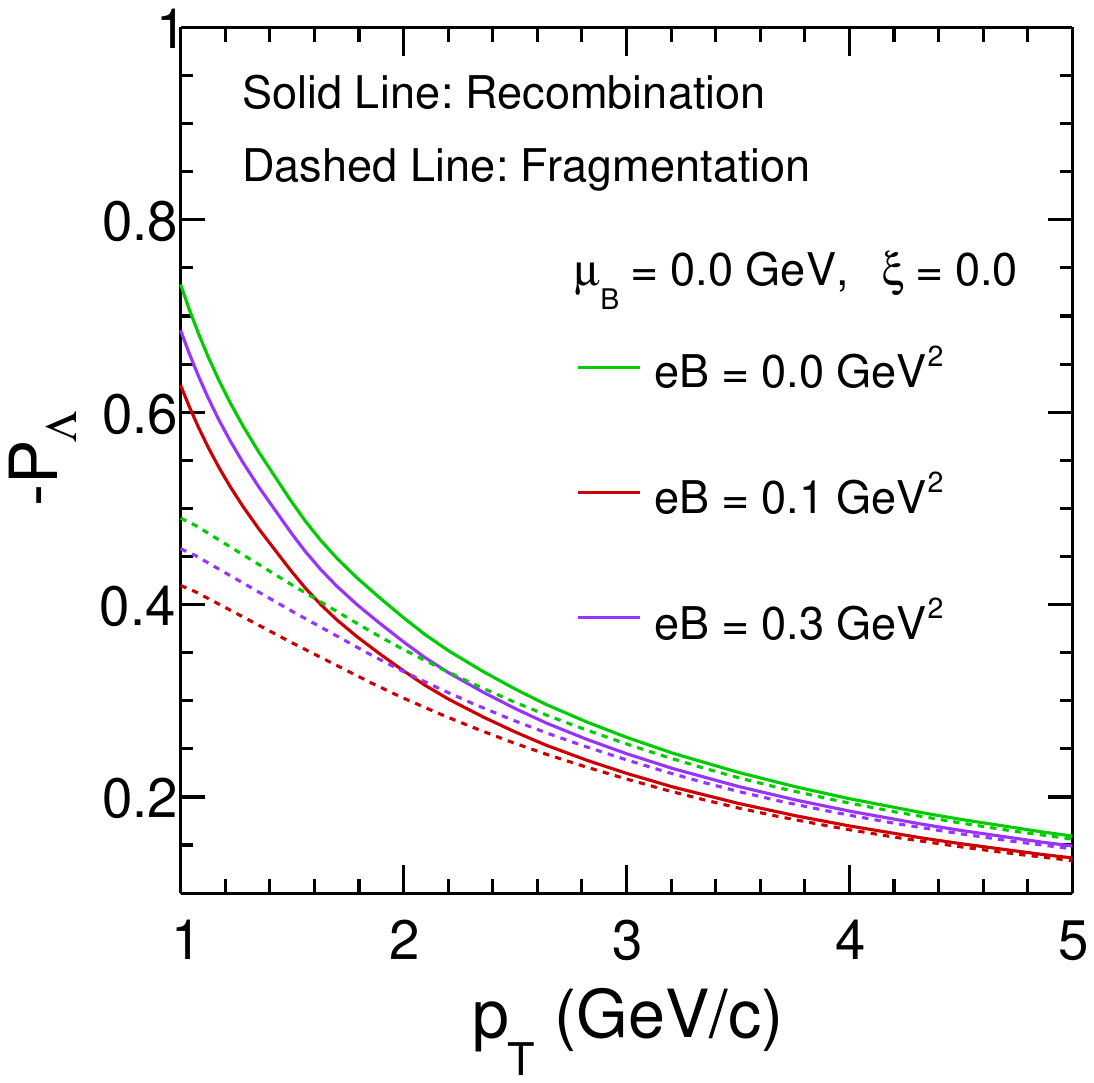}     
\includegraphics[scale=0.29]{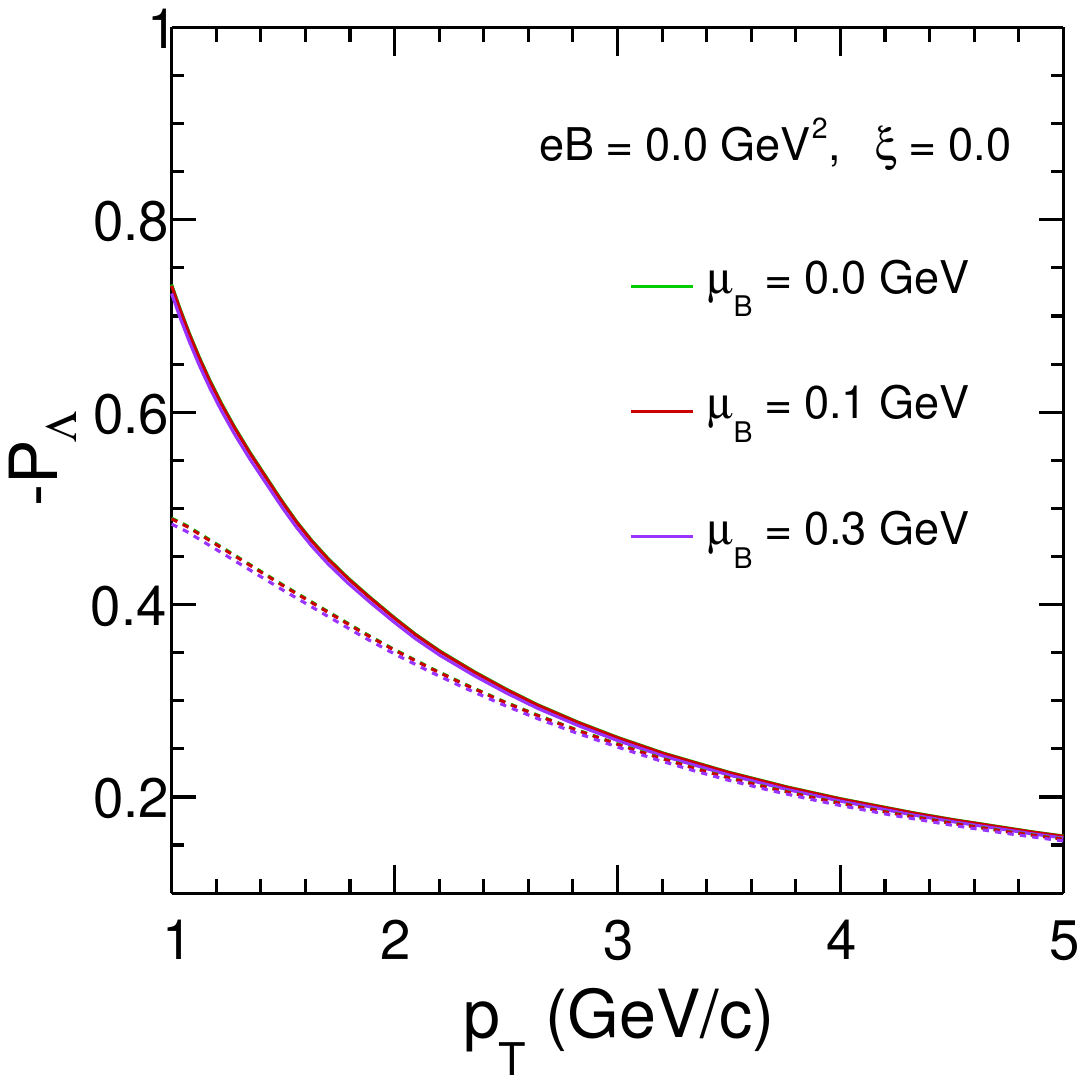}
\includegraphics[scale=0.29]{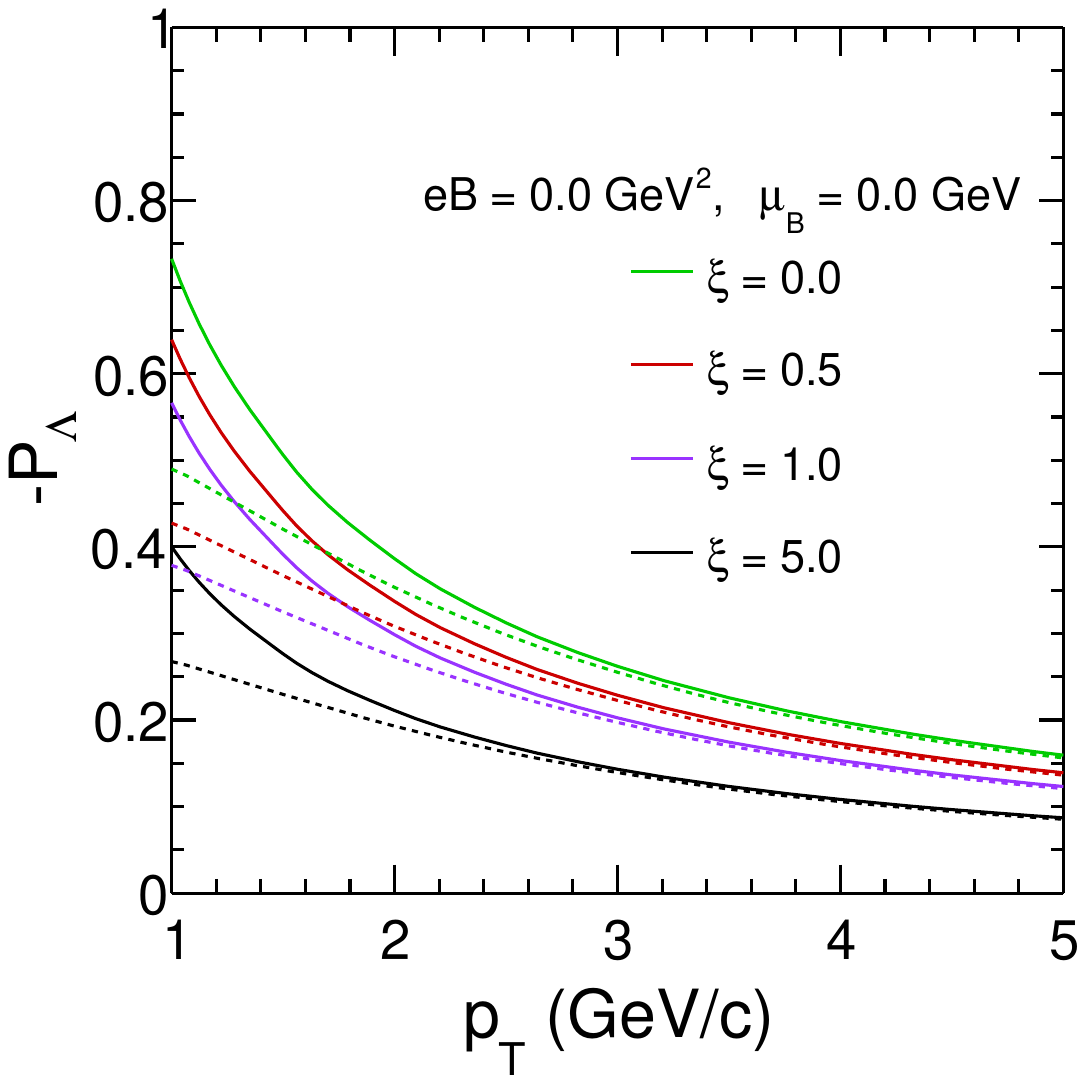}

\caption{(Color online) The variation of $\Lambda$ hyperon polarization ($P_{\Lambda}$) as a function of transverse momentum ($p_{\rm T}$) due to recombination (solid line) and the fragmentation (dashed line) process of hadronization in the presence of magnetic field ($eB$) (left panel), baryon chemical potential ($\mu_{B}$) (middle panel), and anisotropic 
parameter ($\xi$) (right panel) at temperature T = 200 MeV. }
\label{fig:2}
\ec
\end{figure*}

The left panel of Fig.~\ref{fig:1} shows that the Debye mass increases with magnetic field and temperature. However, the impact of 
the external magnetic field on the rate of change in Debye mass is stronger at lower temperatures. The observed enhancement in the Debye 
mass with the external magnetic field is in qualitative agreement with the findings reported in Refs.~\cite{Bonati:2017uvz, 
Bandyopadhyay:2016fyd, Bandyopadhyay:2019pml}. The middle panel of Fig.~\ref{fig:1} reveals a modest decrease in the Debye mass with 
increasing baryon chemical potential in a baryon-rich medium. Although its effect is minimal in magnitude, it is a bit prominent at lower temperatures compared to higher temperatures. In the right panel of 
Fig.~\ref{fig:1}, we found that the Debye mass decreases significantly as the anisotropy parameter increases. The decrease in the Debye mass with the momentum anisotropy agrees well with the finding reported in Refs.~\cite{Dong:2021gnb, Dong:2022mbo}. \\

Based on these observations, it can be concluded that the medium effects are encoded in the Debye mass. It is more sensitive to the 
anisotropy parameter than the magnetic field and baryon chemical potential. The  Debye mass of the medium represents the change in 
the degree of freedom of the systems. If the Debye mass of the system is small, then the color degrees of freedom would be reduced 
to the hadron degree of freedom. As the Debye mass depends on the medium temperature (energy) and the coupling strength 
of the medium. The magnetic field in the medium boosts the energy level and strengthens the coupling constant. 
Due to this, a positive change in Debye mass with increasing the magnetic field can be seen in the left panel of Fig.~\ref{fig:1}. Conversely, finite baryon chemical potential also provides additional energy to the medium but reduces the 
coupling strengths. Consequently, in Fig.~\ref{fig:1}, we observe the decrease in the Debye mass with increasing baryon 
chemical potential. Meanwhile, the anisotropy of the medium does not affect the coupling strength of the medium. However, an uneven 
distribution of the partons in the medium measured in terms of $\xi$ reduces the effective mass. Therefore, a 
visible reduction in the Debye mass with large values of $\xi$ is observed in Fig.~\ref{fig:1}. The combined effect of the magnetic 
field, baryon chemical potential, and medium anisotropy on Debye mass can be estimated using Eq.~\ref{eq15}, which is not explicitly 
illustrated in this work. But one can infer the combined effect of these parameters on Debye mass from Fig.~\ref{fig:1}. \\

\begin{figure*}[!htp]
\bc   
\includegraphics[scale=0.29]{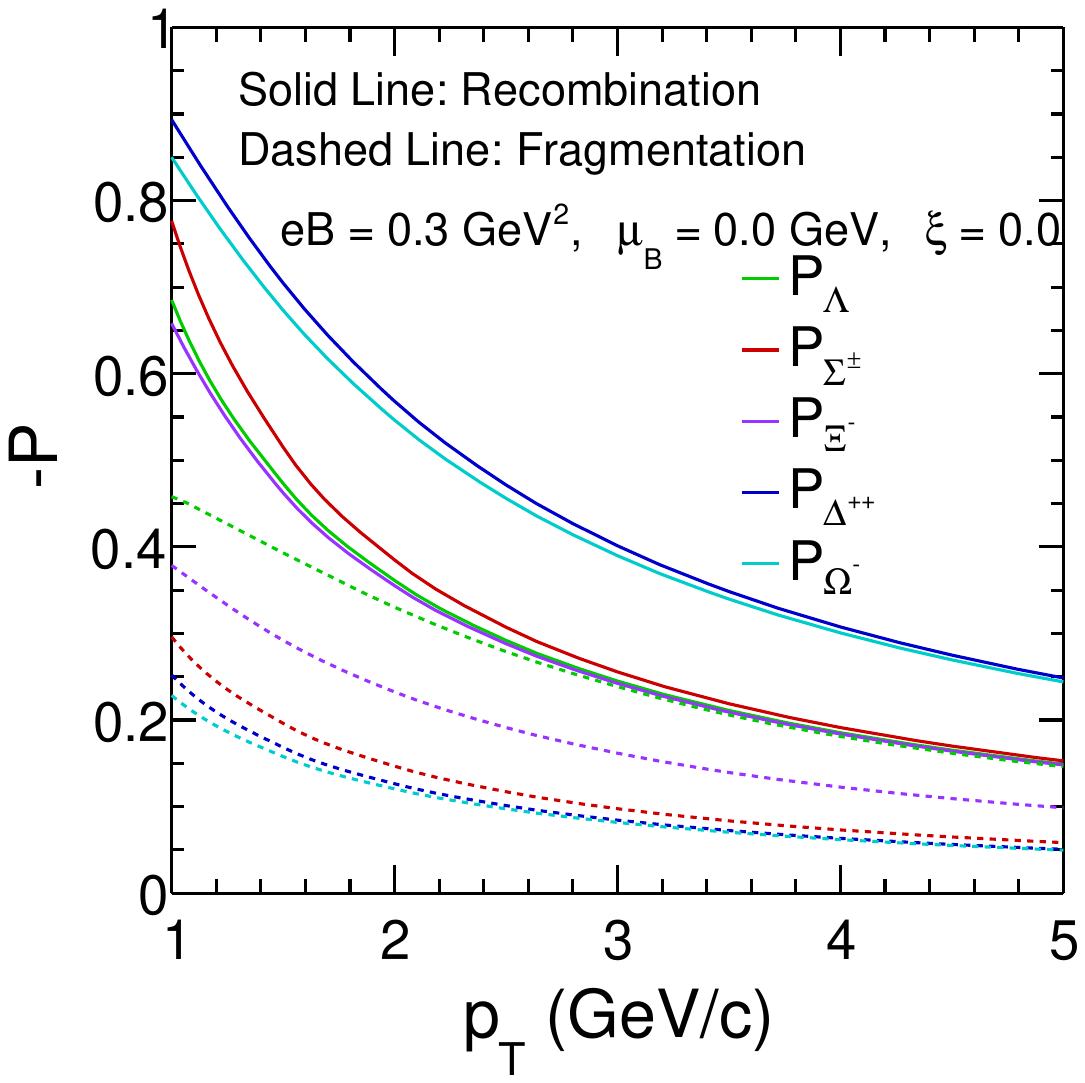}     
\includegraphics[scale=0.29]{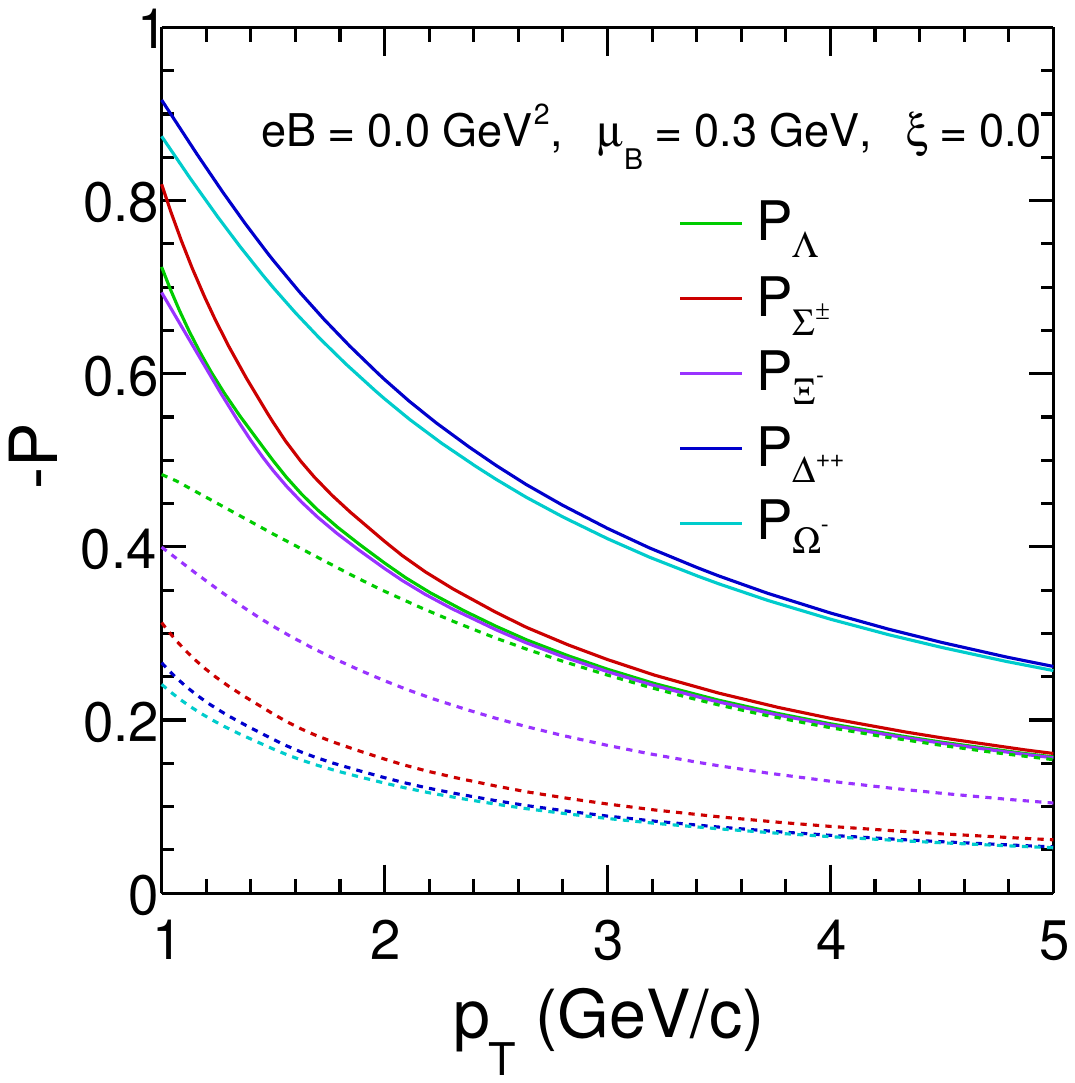}
\includegraphics[scale=0.29]{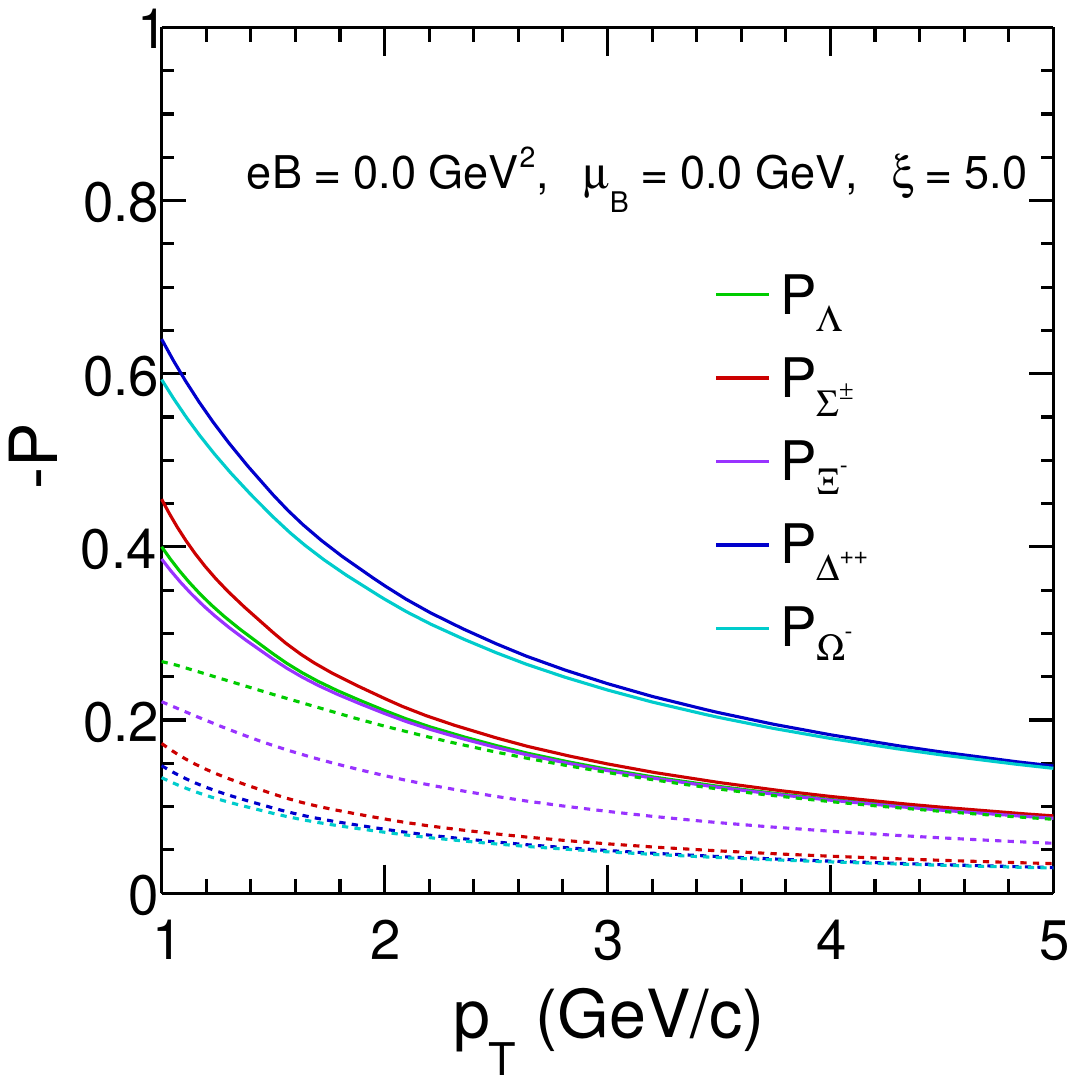}

\caption{(Color online) The variation of baryon polarization as a function of transverse momentum ($p_{\rm T}$) in the recombination (solid line) and fragmentation (dashed line) process of hadronization due to magnetic field ($eB$) (left panel), baryon chemical potential ($\mu_{B}$) (middle panel), and anisotropic parameter ($\xi$) (right panel) at temperature T = 200 MeV.}

\label{fig:3}
\ec
\end{figure*}

\subsection{Spin polarization of hyperons}
\label{baryon}
It is crucial to understand how these polarized quarks recombine to produce the final polarized hyperons. The spin polarization of hyperons depends on the process of hadronization. The spin polarization of $\Lambda$, $\Sigma^{\pm}$, $\Xi^{-}$, $\Delta^{++}$, and $\Omega^{-}$ hyperons through quark recombination process is obtained using the Eqs.~\ref{PolLambdarec},~\ref{Polsigmarec},~\ref{Poldeltarec} and fragmentation process is obtained using Eqs.~\ref{Pollambdafrag},~\ref{Polcascadefrag},~\ref{PolOmegafrag}.  The solid and dashed lines in Fig.~\ref{fig:2} depict the change in the $\Lambda$ spin polarization with $p_{T}$ corresponding to the recombination and fragmentation process of hadronization, respectively. The spin polarization corresponding to the fragmentation process is obtained using the $n_s$ and $f_s$ parameters. The values of these parameters are extracted from Ref.~\cite{Zhang:2018vyr}. The left panel of Fig.~\ref{fig:2} shows the variation of $\Lambda$ hyperon spin polarization as a function of the transverse momentum ($p_{T}$) with the magnetic field at $\mu_{B}$ = 0 GeV and $\xi$ = 0. We consider two finite values of the magnetic field, i.e., eB = 0.1, 0.3 GeV$^2$ to study the effect of a strong magnetic field on spin polarization. The choice of magnetic field in the present study is purely theoretical. The magnetic field of order $eB \approx$ 0.3 GeV$^2$ is used to investigate the impact of a higher magnetic field on the spin polarization observable. Here, we consider the lowest Landau level (LLL) approximation under the strong magnetic field limit while calculating the quark polarization. The left panel of Fig.~\ref{fig:2} indicates the spin polarization of $\Lambda$ hyperon decreases as a function of the $p_{T}$ for considered values of the magnetic field in both processes of the hadronization mechanism. The $p_{T}$ dependence of $\Lambda$ hyperon spin polarization in the presence of an external magnetic field in the fragmentation process follows the same trend as the recombination process, except a change in the slope is observed at low $p_{T}$. However, the magnitude of $\Lambda$ spin polarization due to fragmentation is comparably less than the recombination process. It is observed that the increase in the spin polarization of the $\Lambda$ hyperon with a magnetic field can be attributed to the corresponding increase in the Debye mass. Apart from that, the magnetic field helps to polarize the particle and anti-particle through their magnetic moments. The $\Lambda$ hyperon has a negative magnetic moment, whereas the $\bar{\Lambda}$ hyperon has a positive magnetic moment. Therefore, in the presence of a magnetic field, the spin of the $\bar{\Lambda}$ is aligned along the direction of the field, while the $\Lambda$ spin is aligned against the direction of the field. Therefore, earlier it was suggested that magnetic-field could be one of the possible reasons to describe the splitting of $\Lambda$ and $\bar{\Lambda}$ hyperon data qualitatively with $P_{\bar{\Lambda}} > P_{\Lambda}$~\cite{Becattini:2016gvu, Han:2017hdi, Xu:2022hql, Guo:2019mgh, Guo:2019joy}. However, various recent studies suggest the total 6\% polarization splitting of $\Lambda$ and $\bar{\Lambda}$ hyperon at $\sqrt{s_{NN}}$ = 7.7 GeV remains unexplained by a single mechanism so far~\cite{STAR:2017ckg}, rather need for a combination of different mechanisms. For example, in Ref.~\cite{Ryu:2021lnx}, the authors have argued the gradient of $\mu_{B}$/T can alter the ordering in the spin polarization of $\Lambda$ and $\bar{\Lambda}$ hyperons. In Ref.~\cite{Wu:2022mkr}, the calculation shows that the distinctive effects of baryon chemical potential and the production time of $\Lambda$ and $\bar{\Lambda}$ hyperons predict a small value of polarization splitting.  Further, in Ref.~\cite{Wu:2023dfz}, the authors have explicitly estimated the contribution to the observed polarization splitting due to three mechanisms; such as the meson field mechanism~\cite{Csernai:2018yok}, different freeze-out space-time mechanism~\cite{Vitiuk:2019rfv}, and the due to magnetic field~\cite{Guo:2019mgh} by using the particle-in-cell relativistic (PICR) hydrodynamic model. These three different mechanisms lie in different stages of high-energy collisions. Thus, they are not contradicted by each other, and the meson field mechanism is found to be dominant. However, recently with the better precision measurement in the Beam Energy Scan-II data of the STAR Collaboration measured no significant splitting between the $\Lambda$ and $\bar{\Lambda}$ hyperon at $\sqrt{s_{NN}}$ = 19.6  and 27 GeV~\cite{STAR:2023nvo}. The current formalism does not necessarily differentiate between particle and anti-particle and, therefore, predicts the identical spin polarization for $\Lambda$ and $\bar{\Lambda}$ hyperons.

During the exclusive recombination process of hadronization, it has been observed that the spin polarization of $\Lambda$ hyperons aligns with the spin polarization of the initial strange quark. The initial strange quarks with lower momentum have higher values of $\Lambda$ spin polarization after recombining to $\Lambda$ hyperons and vice-versa. The middle panel of Fig.~\ref{fig:2} illustrates the dependence of $\Lambda$ hyperon spin polarization as a function of $p_{T}$ for various values of chemical potential in the recombination and fragmentation scenario of hadronization marked in the solid and dashed line, respectively. Two finite values of chemical potential are considered with $\mu_{B}$ = 0.1, 0.3 GeV. It can be seen here that there is a very marginal effect of baryon chemical potential on the spin polarization of $\Lambda$, which is consistent with the corresponding Debye mass result shown in Fig.~\ref{fig:1}. With the considered values of $\mu_{B}$, the temperature-dependent term in Eq.~\ref{mDmuB} dominates over the $\mu_{B}$-dependent terms hence giving insignificant effect of baryon chemical potential. The right panel of Fig.~\ref{fig:2} depicts the variation of $\Lambda$ spin polarization as a function of $p_{T}$ for various values of the anisotropic parameter. Anisotropy in the medium allows the alignment of particle spins along a specific direction. Here, we considered four different values of anisotropic parameters, i.e., $\xi$ = 0.5, 1.0, 5.0, along with the isotropic case, i.e., $\xi$ = 0. It is observed that the $\Lambda$ hyperon spin polarization decreases with the anisotropy parameter. It is an intriguing observation that the medium anisotropy plays a significant role in influencing the spin polarization of $\Lambda$ hyperons.\\

Furthermore, in Fig.~\ref{fig:3} the variation of $\Lambda$, $\Sigma^{\pm}$, $\Xi^{-}$, $\Delta^{++}$, 
and $\Omega^{-}$ hyperons spin polarization is shown as a function of $p_{T}$ due to the recombination 
and fragmentation process of hadronization in the solid and dashed line, respectively. The left panel of 
Fig.~\ref{fig:3} depicts the spin polarization of these hyperons at $eB = 0.3$ GeV$^{2}$, $\mu_{B}$ = 0 
GeV and $\xi$ = 0. The middle and right panel of Fig.~\ref{fig:3} shows $\Lambda$, $\Sigma^{\pm}$, 
$\Xi^{-}$, $\Delta^{++}$, and $\Omega^{-}$ spin polarization with $p_{T}$ for baryon chemical potential 
$\mu_{B} = 0.3$ GeV and anisotropic parameter $\xi$ = 5, respectively. The role of baryon chemical 
potential on particle spin polarization is quite similar to the magnetic field. Moreover, we observe 
that the spin polarization of all the considered hyperons is dominant at low $p_T$ in the recombination 
process. However, in the fragmentation process, the spin polarization of $\Delta^{++}$ and $\Omega^{-}$  
is  1/3 of the spin polarization due to the u and s quarks, respectively. Therefore, baryons having the 
same quark flavor, each quark contributes equally to the final state baryon polarization. Meanwhile, 
baryons of mixed quark flavors have linear combinations due to each flavor. As consequence, a 
significant decrease in the $\Delta^{++}$ and $\Omega^{-}$ spin polarization is observed compared to 
$\Lambda$, $\Sigma^{\pm}$, and $\Xi^{-}$ as shown in Fig.~\ref{fig:3}. The change in the spin polarization due to 
fragmentation in the presence of the magnetic field, baryon chemical potential, and medium anisotropy 
follows the same trend of the spin polarization obtained from the recombination process.  However, the 
spin polarization in the fragmentation process is always less compared to recombination for all baryons. 
Fig.~\ref{fig:3} indicates a quark flavor-dependent spin polarization of hyperons in the heavy-ion 
collisions. We found that there is grouping among the particles of baryon octet with spin-1/2 and baryon 
decouplet with spin-3/2 during the spin-orbit coupling. Particles having the same quark flavor, for 
example, $\Omega^{-}$ (sss) and $\Delta^{++}$ (uuu)  have similar spin-orientation along the direction 
of OAM as compared to particles having mixed quark flavor, e.g., $\Lambda$ (uds), $\Sigma^{\pm}$ 
(uus/dds), and $\Xi^{-}$ (dss). Therefore, hadrons with the same flavor possibly have some correlation 
among themselves, which drives the spin polarization. This correlation might be absent in the hadron with different 
flavors; such an observation is recently reported for $\phi$ meson in Ref.~\cite{Sheng:2022wsy}. In 
addition, the baryon-decouplet of spin-3/2 has higher spin polarization as compared to the baryon-octet of 
spin-1/2. Therefore, particle spin might play a role in its polarization. The recent measurement global 
spin polarization of $\Xi$ and $\Omega$ hyperons at STAR hint at a possible hierarchy in the global spin 
polarization, i.e., $P_{\Lambda} < P_{\Xi} < P_{\Omega}$ ~\cite{STAR:2020xbm}. The large systematic and 
statistical uncertainties raise a question as to whether the difference in spin polarization of various 
hadrons is either due to the mass, lifetime, strangeness quantum number, differential freeze-out, and/or 
any localized effects, etc. 

In literature, it is derived that at the local thermodynamic equilibrium, the 
mean spin vector of fermions at the leading order of thermal vorticity is directly proportional to its 
spin, but inversely proportional to its mass~\cite{Becattini:2013fla, Fang:2016vpj, Becattini:2016gvu}. So, the lighter and higher spin particles are expected to be polarized easily by the fluid vorticity. 
Using the same formalism in a multiphase transport (AMPT) model the spin polarization of $\Lambda$, $\Xi^{0}$ 
and $\Omega^{-}$ hyperons are explored by considering the fluid thermal vorticity into 
account~\cite{Wei:2018zfb}. In the AMPT model calculation, the global spin polarization of single and multi-strange hyperons
shows an ordering on the spin polarization observable such that; $P_{\Xi} < P_{\Lambda} < P_{\Omega}$, in the absence of feed-down effect at the lower center of mass 
energies~\cite{Wei:2018zfb}. However, it is predicted that considering the feed-down effect will suppress primary $\Lambda$ polarization by (15-20)\%, while $\Xi$ hyperons have less contribution due to the feed-down~\cite{Becattini:2016gvu, Karpenko:2016jyx, Li:2017slc, Jiang:2016woz, Kolomeitsev:2018svb, Xia:2019fjf}. This indicates that the multi-strange hadrons may freeze-out at earlier in times and may lead to larger polarization of $\Omega^{-}$ and $\Xi^{0}$ compared to single 
strange $\Lambda$~\cite{Vitiuk:2019rfv}. Similarly, in Ref.~\cite{Voronyuk:2023vyu}, the authors have 
studied the spin polarization of several baryons such as; $\Lambda$($\bar{\Lambda}$), $\Xi^{\pm}$, 
$\Xi^{0}$($\bar{\Xi^{0}}$), $\Sigma^{0}$($\bar{\Sigma}^{0}$), and $\Omega$($\bar{\Omega}$) hyperons induced by the local vorticity in a parton-hadron-string dynamic (PHSD) model at lower energies between $\sqrt{s_{NN}}$ = 3.5 GeV to 11.5 GeV. With the inclusion of the feed-down effect, an ordering in the hyperon spin polarization is found such that; $P_{\bar{\Xi}} \simeq P_{\bar{\Lambda}} > P_{\bar{\Sigma^{0}}} > P_{\Lambda} > P_{\Sigma^{0}} > P_{\Xi}$. But the behavior of $\Omega$($\bar{\Omega}$) polarization with the center of mass energy seems to disagree with the energy trend of other hyperons. Although global spin polarization for different hyperons is found to be different, the qualitative agreement of the hydrodynamical prediction with experimental data suggests the underlying origin of spin polarization could be dominated by hydrodynamic fields~\cite{Vitiuk:2019rfv, Wei:2018zfb}. It can be inferred that spin polarization is predominantly driven by the collective phenomena of the system rather than the specific flavor-dependent coupling~\cite{Becattini:2024uha}. Furthermore, to signify the role of spin and flavor dependence in the global polarization and to better understand the spin polarization mechanism in heavy-ion collisions, the experimental verification of global spin polarization for different particles with different spin and/or magnetic moments is required.\\

\begin{figure*}[!htp]
\bc   
\includegraphics[scale=0.42]{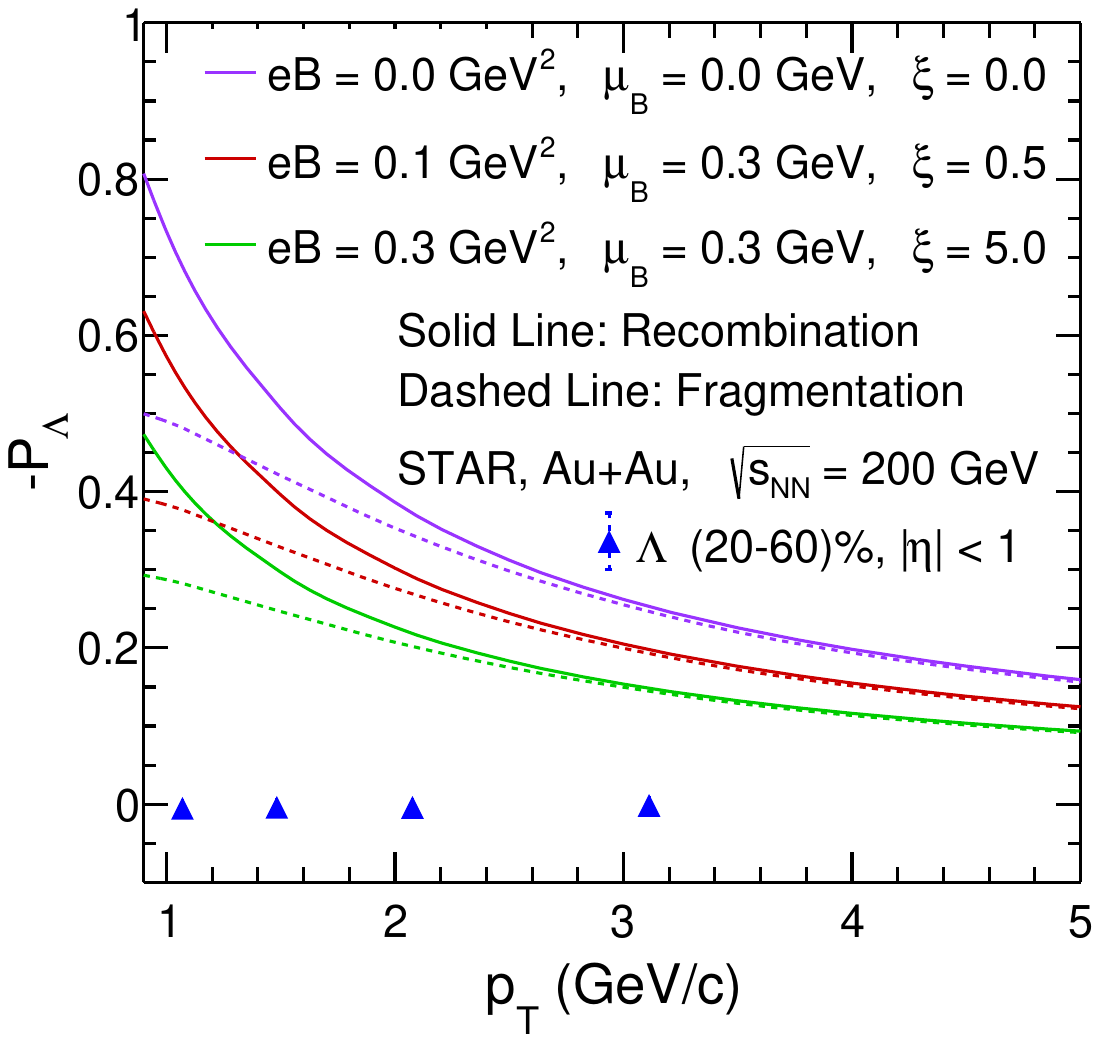}
\includegraphics[scale=0.42]{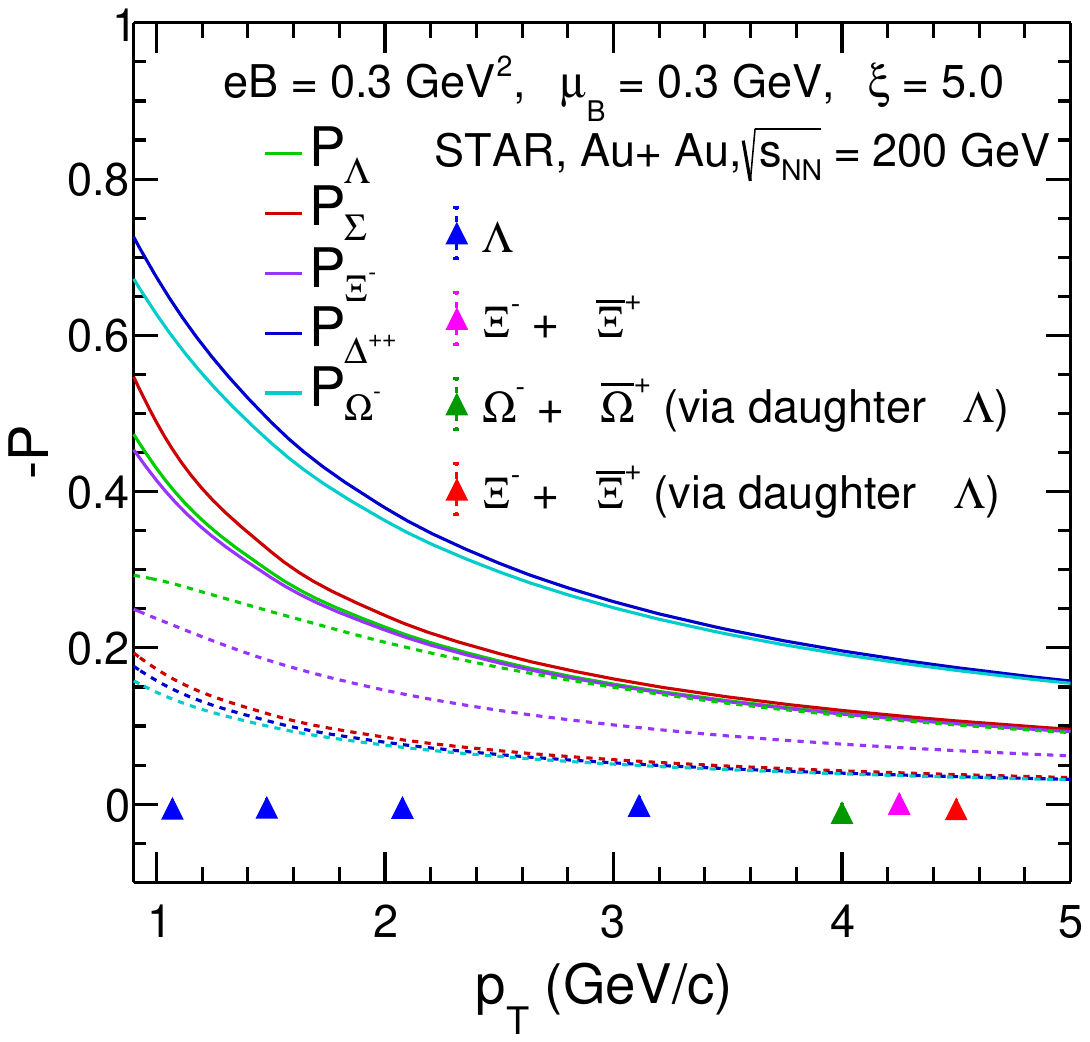}

\caption{(Color online)The variation of $\Lambda$ hyperon (left panel), and
$\Sigma^{\pm}$, $\Xi^{-}$, $\Delta^{++}$, and $\Omega^{-}$ hyperons (right panel) spin
polarization as a function of transverse momentum due to combined effect of magnetic
field, baryon chemical potential, and anisotropic parameter at temperature T = 200 MeV. The results are compared with STAR measurements taken from 
Ref.~\cite{STAR:2018gyt, STAR:2020xbm}. The statistical and systematic uncertainties of the experimental 
data points are within the marker size. It should be noted here that the data points for $\Xi$ and $\Omega^{-}$ are $p_{\rm T}$-integrated. However, for a comparison we have put those data on the same plot with $p_{\rm T}$ around 4 GeV/c.}
\label{fig:4}
\ec
\end{figure*}

So far, we have discussed the effect of magnetic field, baryon chemical potential, and medium anisotropy 
on the spin polarization of hyperons separately. Now, the combined effect of these three parameters is 
examined in Fig.~\ref{fig:4}. The left panel of Fig.~\ref{fig:4} shows the combined effect on $\Lambda$ 
hyperon spin polarization.  The obtained results are compared with STAR experimental data of $\Lambda$ 
hyperons for Au+Au collisions at $\sqrt{s_{NN}}$ = 200 GeV at (20-60)\% 
centrality class~\cite{STAR:2018gyt}. The comparison with experimental data quantifies the contribution of Debye mass on hadron spin polarization in relativistic heavy-ion collisions. It is noteworthy to mention that the experimental data comparison lacks one-to-one correspondence because in the theoretical estimation the direct correspondence of the detector acceptance, collision 
geometry, center of mass energy, event plane resolution, feed-down corrections, etc., are not considered into account.
The current study predicts a comparably larger global spin polarization than the 
experimental data for $\Lambda$ hyperon as the function of $p_{T}$. The possible reason for this difference is that the global spin polarization obtained in this study considered only the change in the medium Debye screening mass. In literature, it is proposed that the hyperon spin polarization may arise due to several other sources such as thermal vorticity, thermal shear, gradient of baryon chemical potential over temperature (also called the spin Hall effect), fluid acceleration, electromagnetic fields, etc.~\cite{Wu:2022mkr, Yi:2021unq}. The results obtained in this study are certainly considered more likely a preliminary estimation rather than a quantitative analysis because of the above-mentioned reason. However, the spin polarization probability due to Debye mass is independent of the aforementioned polarization mechanisms. Combining all the possible sources for global spin polarization along with Debye mass-based polarization could possibly lead to spin polarization comparable with experimental data. \\

The right panel of Fig.~\ref{fig:4} shows 
change in the spin polarization of $\Lambda$, $\Sigma^{\pm}$, $\Xi^{-}$, $\Omega^{-}$, and $\Delta^{++}$ 
baryons due to the combined effect of $eB$, $\mu_{B}$ and $\xi$. In the right panel of Fig.~\ref{fig:4}, the obtained results are compared with the experimental data of $\Lambda$ hyperons and the averaged spin polarization of $\Xi$ and $\Omega$ hyperons measured for Au+Au collisions at  $\sqrt{s_{NN}}$ = 200 GeV at (20-80)\% centrality class~\cite{STAR:2020xbm}. The global spin polarization of $\Xi$ hyperons was measured directly and via polarization transfer method to decay daughters, while for $\Omega$ hyperons, it is obtained via transfer to decay daughters~\cite{STAR:2020xbm}. Due to the unavailability of $p_{T}$ differential $\Xi$ and $\Omega$ hyperons global spin polarization data, we have compared to the $p_{T}$ integrated value. The medium anisotropy has a prevalent impact on the spin polarization observable compared to the magnetic field and baryon chemical potential. The observed feature of medium anisotropy on polarization is due to the influence of anisotropy on Debye mass. Medium anisotropy plays a crucial role in controlling the spin polarization of hyperon, which needs an extensive investigation from a theoretical standpoint. \\

\begin{figure*}[htp]
\bc    
\includegraphics[scale=0.42]{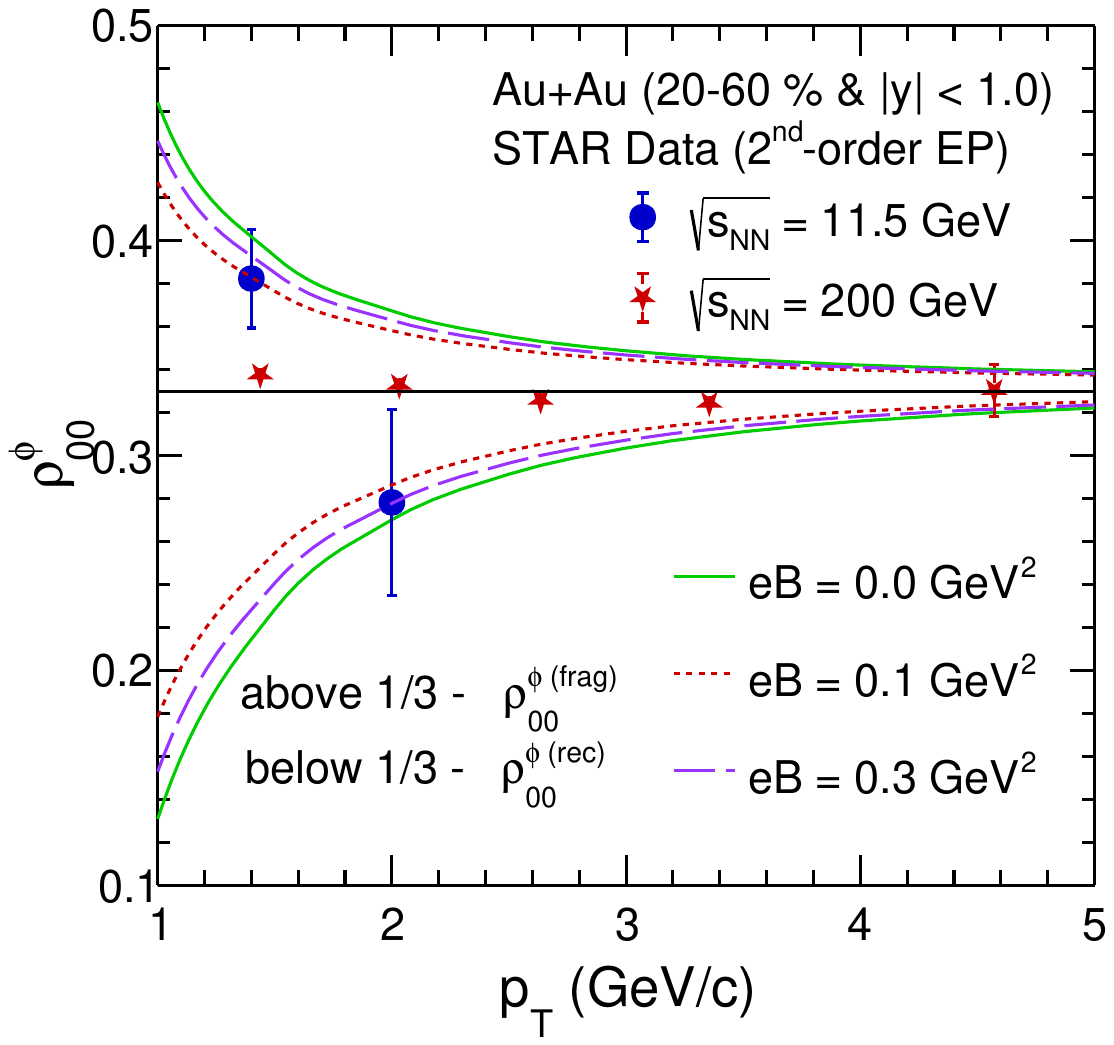}
\includegraphics[scale=0.42]{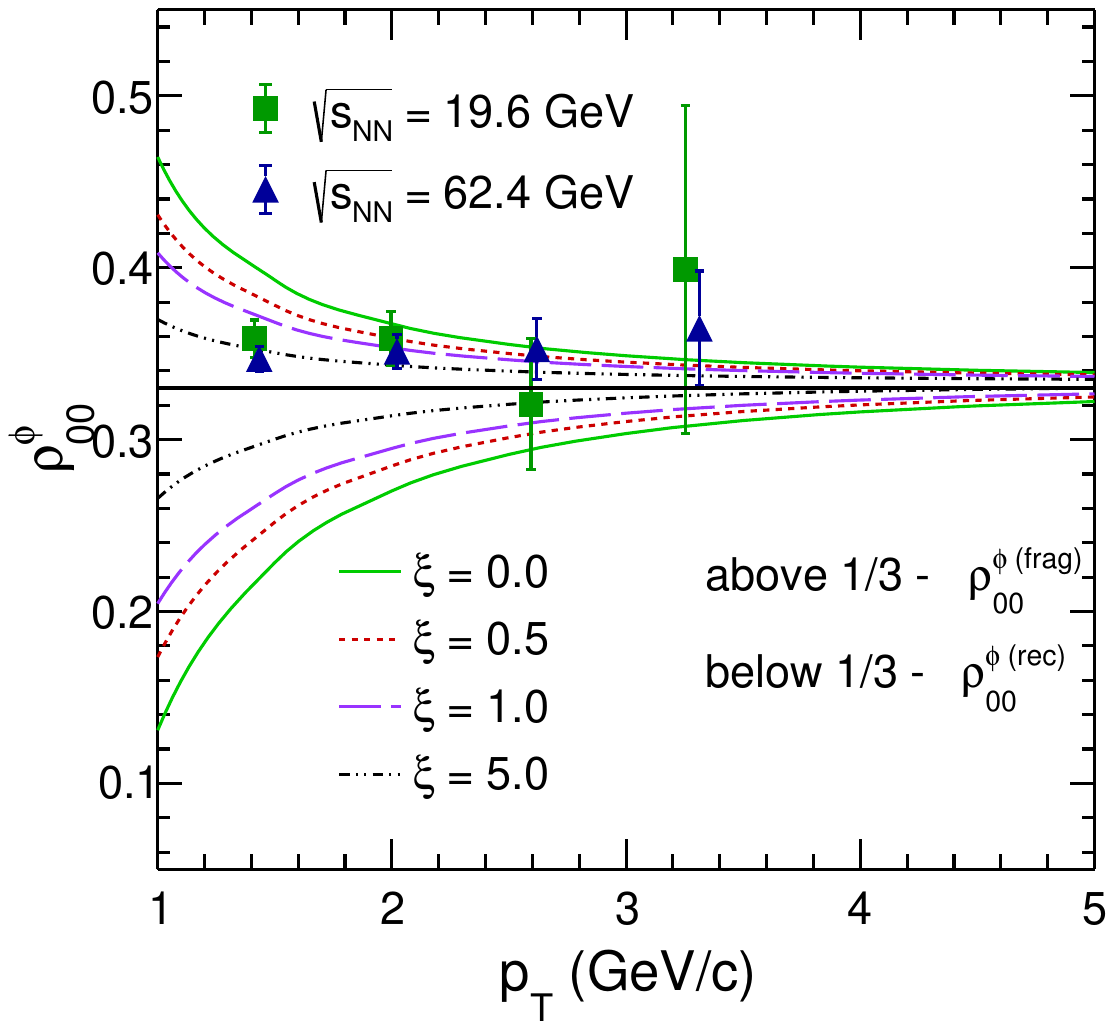}
\caption{(Color online) The variation of spin alignment of $ \phi $ vector meson as a function of transverse momentum due to the magnetic field (left panel) and anisotropic parameter (right panel) at temperature T = 200 MeV. The results are compared with STAR measurements taken from Ref.~\cite{STAR:2022fan}.}

\label{fig:5}
\ec
\end{figure*}    

\begin{figure*}[!htp]
\bc   
\includegraphics[scale=0.42]{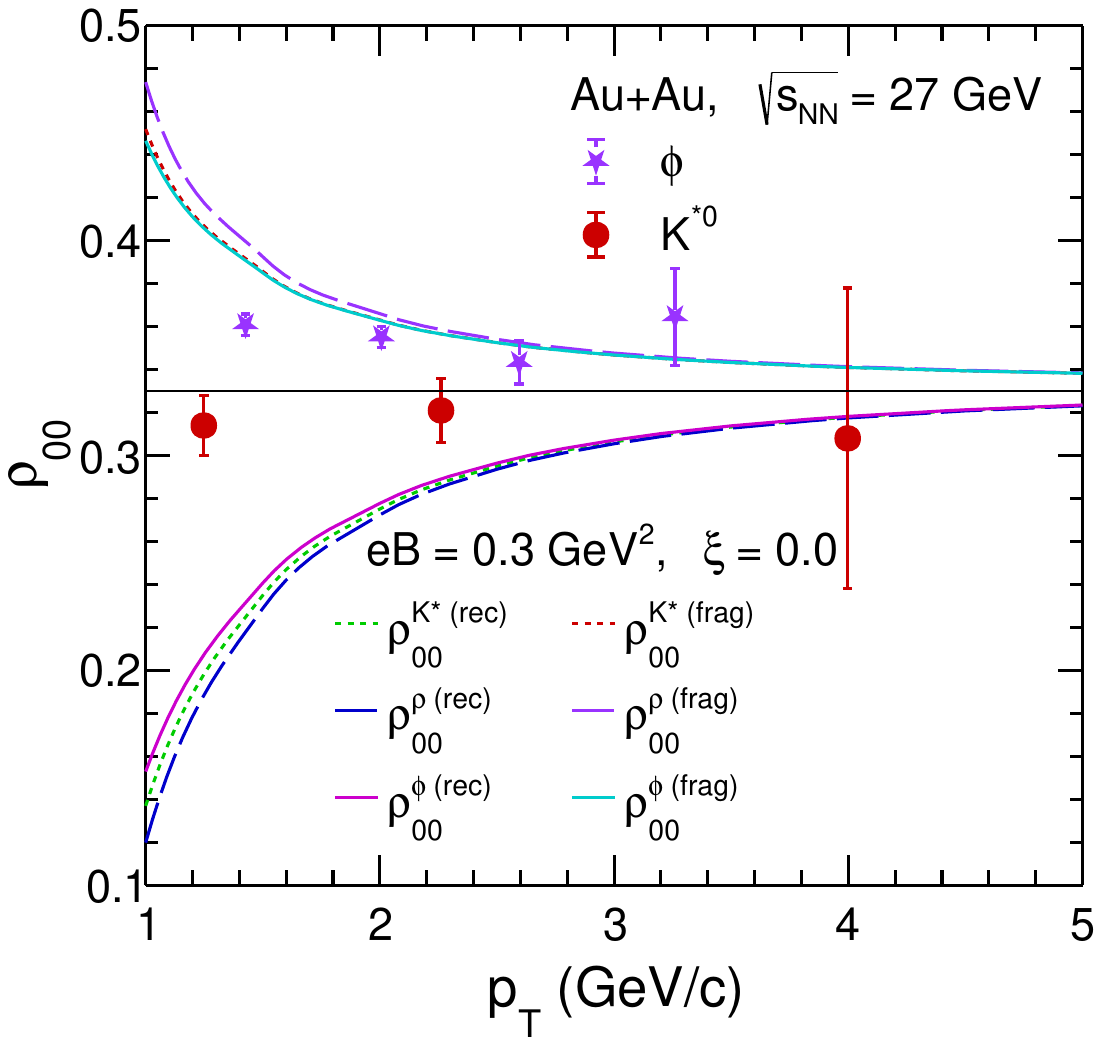}   
\includegraphics[scale=0.42]{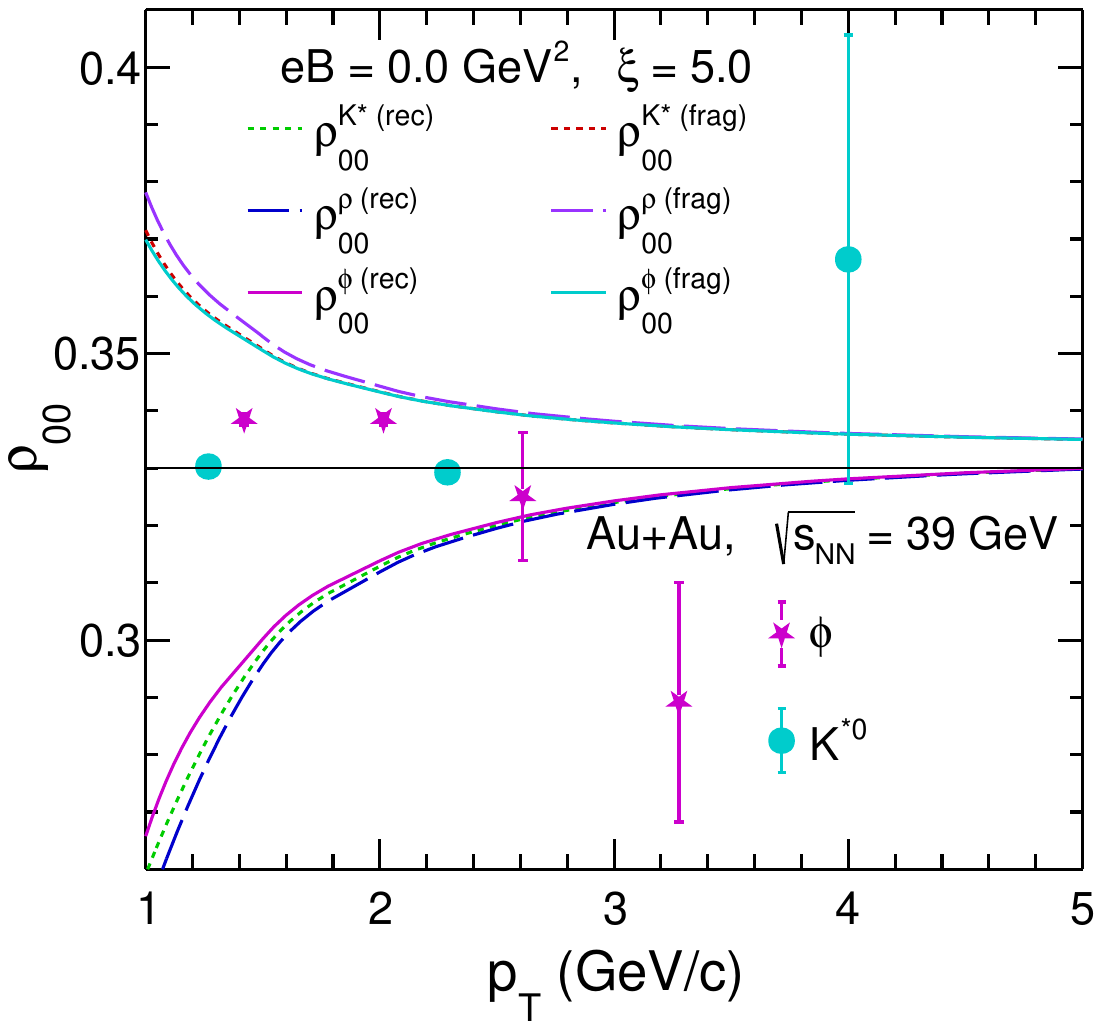}
\caption{(Color online) The variation of spin alignment of $ \phi $, $K^{*0}$, $\rho$ vector meson as a function of transverse momentum due to the magnetic field (left panel), and anisotropic parameter (right panel)  at temperature T = 200 MeV. The results are compared with STAR measurements taken from Ref.~\cite{STAR:2022fan}.}

\label{fig:6}
\ec
\end{figure*}

\subsection{Spin alignment of vector mesons }
\label{vectormeson}

Similar to the global spin polarization of hyperon, the spin alignment of vector meson is also based on the same mechanism of spin-orbit 
coupling. Since a vector meson with spin one can have three different spin orientations, the probability for its spin to align in a given 
direction is 1/3. Any value of the spin alignment differs from 1/3, which means the polarization of the vector meson is along that direction. With 
the recent large spin alignment measurement of $\phi$ vector meson at  RHIC~\cite{STAR:2022fan, Singha:2020qns} and 
LHC~\cite{ALICE:2019aid}, it becomes a challenging and active area of research for the theoretical community to study the possible 
sources of spin alignment for $\phi$ vector meson. The spin alignment of vector meson comes from both constituent quark and anti-quark 
and hence its spin polarization is quadratic in  vorticity~\cite{Yang:2017sdk}.\\

The left panel of Fig.~\ref{fig:5} shows the effect of magnetic field on the spin alignment of $\phi$ meson for two different 
hadronization scenarios as described in formalism section~\ref{formulation}. The curves below line 1/3 depict the spin alignment of 
$\phi$-meson due to the recombination of polarized quarks and anti-quarks. The curves above line 1/3 show the hadronization process of 
quark and anti-quark pairs in which either of them is created in the fragmentation process. We compare the obtained results with the 
STAR Collaboration data for Au+Au collisions for various centers of mass energies in (20-60)\% class~\cite{STAR:2022fan}\footnote{The choice of 
center of mass energy selection of STAR data for Au+Au collisions in Fig.~\ref{fig:5} and Fig.~\ref{fig:6} is just for illustration 
purposes, and it has nothing specific relation with the magnetic field and anisotropy chosen in this study.}.  It is found that the absolute deviation of $\rho_{00}$ from 1/3 ($ | \rho_{00} - 1/3|$) for 
$\phi$ vector meson increases with magnetic field for both cases of hadronization processes. 
The magnetic field aligns the particle spin 
in a specific direction; hence, it helps to enhance the degree of spin alignment corresponding the the non-zero magnetic field.  The right panel of Fig.~\ref{fig:5} shows the effect of $\xi$ on the spin alignment of $\phi$ meson with a set of $\xi$ values. From the figure, it is evident that medium anisotropy significantly modifies the vector meson spin alignment. The left panel of Fig.~\ref{fig:3} shows the  spin alignment of $\phi$, $K^{*}$, and $\rho$ meson at $eB=0.3$ GeV$^{2}$ for corresponding to fragmentation and recombination hadronization scenarios.  The right panel of Fig.~\ref{fig:4} shows the spin alignment of $\phi$, $K^{*}$, and $\rho$ meson for $\xi$ = 5. Figure~\ref{fig:5} and ~\ref{fig:4} depicts the absolute deviation of $\rho_{00}$ from 1/3 is more at low $p_{T}$ for the recombination process as compared to the fragmentation. This implies that the degree of spin alignment is more at low $p_{T}$ in the recombination process for $\phi$, $K^{*}$, and $\rho$ vector mesons. The value of $\rho_{00}$ is close to 1/3 towards high $p_{T}$ showing zero spin alignment for $\phi$,  $K^{*}$, and $\rho$ mesons. In the quark polarization model, the polarization of a quark is inversely proportional to the square of its mass, which suggests a mass ordering in the spin alignment effect of vector mesons due to their constituent quark composition. Figure~\ref{fig:4} depicts $\rho_{00}^{\rho} > \rho_{00}^{K^{*}} > \rho_{00}^{\phi}$ at low $p_T$ in the recombination scenario.

\section{Conclusion}
\label{sum}

In summary, we explored the global spin polarization of hyperons and spin alignment of vector mesons in the presence of a magnetic field, baryon chemical potential, and medium momentum-space anisotropy. The individual and combined effect of these parameters on hyperons and vector mesons spin polarization have been studied in the recombination and fragmentation process of hadronization. We found that the spin polarization of hyperons and spin alignment of vector mesons increases with the increase in magnetic field. In this study, the baryon chemical potential has negligible dependence on hyperon spin polarization. The medium anisotropy significantly affects the spin polarization and spin alignment of hadrons. It is interesting to note that the recombination process dominates at low $p_{T}$ in comparison with the fragmentation process. It implies that hadronization mechanisms play a crucial role to estimate the spin polarization of hadrons. Therefore, for a comprehensive study, one may explore the global spin polarization of hadrons considering different hadronization mechanisms in realistic hydrodynamic or transport model simulations.\\

The hyperon spin polarization obtained through the global quark polarization method is comparably higher than the experimentally measured hyperon spin polarization in heavy-ion collisions at mid-rapidity at the RHIC energies.  However, a thermodynamic three fluid dynamic (3FD) model considering total rapidity into account predicts a higher net hyperons spin polarization and it increases with the increasing center of mass energies~\cite{Ivanov:2019ern, Ivanov:2022ble}, which suggest to have comprehensive understanding of hyperon spin polarization phenomena, the total rapidity coverage is essential. In addition to the magnetic field, baryon chemical potential, and medium momentum-space anisotropy effects explored in this work, there may be various other sources that may affect the hyperon spin polarization. Such sources could be medium rotation, hadronic interactions, anisotropic flows in the medium, and fluctuation of other force fields such as vorticity, temperature gradient, shear tensor, strong force field, etc. However, the contribution of these sources on hyperon spin polarization and spin alignment of vector mesons has yet to be studied collectively. Additionally, this Debye mass-based polarization may have a substantial implication on particle production yield and must be explored in future studies in ultra-relativistic heavy-ion collisions. The qualitative prediction of global spin polarization is an open challenge and requires further investigation.\\

\section*{Acknowledgement}
Bhagyarathi Sahoo acknowledges the financial aid from CSIR, Government of India. The authors gratefully acknowledge the DAE-DST, Government of India funding under the mega-science project "Indian Participation in the ALICE experiment at CERN" bearing Project No.SR/MF/PS-02/2021-IITI (E-37123).

\appendix
\section{Global quark polarization in the presence of magnetic field}
\label{PqeB}

Let us consider a quark-quark scattering at a fixed impact parameter mediated by virtual gluon (Fig.~\ref{appe}). \\

\begin{figure}[!htp]
\includegraphics[scale=0.5]{qq.pdf}
\caption{ Quark-quark scattering Feynman diagram mediated by virtual gluon}
\label{appe}
\end{figure}

Suppose we are interested in the polarization of quark $q_1$  after the scattering. We, therefore, average over the spins of initial quarks and sum over the spin of quark $q_2$ in the final state. The differential scattering cross section for the above process in momentum space is given by~\cite{Liang:2004ph, Liang:2004xn, Liang:2007ma, Gao:2007bc};

\begin{equation}
d\sigma =   \frac{c_{qq}}{4F} \sum_{\lambda_1, \lambda_2, \lambda_4} \frac{d^3p_3}{(2\pi)^3 2 E_3} \frac{d^3p_4}{(2\pi)^3 2 E_4} \frac{|S_{fi}(q)|^2}{TV}
\label{cs1}
\end{equation}
where, T and V are the interaction time and volume of the space, $c_{qq}$ = 2/9 is the color factor, $F  = \sqrt{(p_1 \cdot p_2)^2 - m_1^2m_2^2}$  = $E_1 E_2 v_{rel}$\; is called the flux factor, $p_i$ =($E_i$, $\vec{p_i}$) is the four-momentum of the quark $q_i$, $\lambda_i$ is the spin of $q_i$, q = $p_1 -p_3 = p_4 -p_2$ is the four-momentum transfer, $\mathcal{M}_{fi}(q)$ is called the scattering amplitude in the momentum space. \\

The scattering matrix element ($S_{fi}$) in terms of momentum space is given by;
\begin{equation}
S_{fi} = <f|S|i> = \mathcal{M}_{fi}(q)(2\pi)^4 \delta^4(p_3 + p_4 - p_1 -p_2)
\label{Sfi}
\end{equation}

Taking a 2-dimensional Fourier transformation to impact parameter space and writing the scattering amplitude in momentum space, we have
\begin{equation}
S_{fi} = \int \frac{d^2 q_{T}}{(2\pi)^2} \mathcal{M}_{f i}(q) e^{-i(\bf{q_T +p_T}) \cdot x_T} (2\pi)^4 \delta^4(p_3 + p_4 - p_1 -p_2)
\label{Sfi1}
\end{equation}
\begin{equation}
S_{fi}^{*} = \int \frac{d^2 k_{T}}{(2\pi)^2} \mathcal{M}_{f i}^{*}(k) e^{i(\bf{k_T +p_T}) \cdot x_T} (2\pi)^4 \delta^4(p_3 + p_4 - p_1 -p_2)
\label{Sfi2}
\end{equation}
Here, we use $x_T$ to denote the impact parameter of the two scattering quarks to distinguish from impact parameter {\bf{b}} = b $\hat{e_x}$ of the two nuclei. $q_T$ ($k_T$) is the transverse momentum transfer from the virtual gluon to the quark.\\

Substituting the Eqs~\ref{Sfi1} and ~\ref{Sfi2} in Eq.~\ref{cs1}  and writing the differential scattering cross-section expression with respect to impact parameter space ($x_T$), we obtain
\begin{align}
\frac{d^2\sigma}{d^2 x_T} = \frac{c_{qq}}{4F} \int \frac{d^2 q_{T}}{(2\pi)^2} \frac{d^2 k_{T}}{(2\pi)^2}  e^{i(\bf{k_T - q_T}) \cdot x_T} \mathcal{M}_{f i}(q) \mathcal{M}_{f i}^{*}(k) \nonumber \\ \delta^4(p_3 + p_4 - p_1 -p_2) \delta^2(q_T + p_T) \frac{d^3p_3}{2 E_3} \frac{d^3p_4}{2 E_4} 
\label{cs2}
\end{align}

In the presence of magnetic field $eB$ (suppose $\textbf{B} = B \hat{z}$) the single charged particle energy eigen value ($E_{iz}$) is given by 
\begin{equation}
E_{iz} = \sqrt{p_z^{2} + m_{i}^2 + 2l| Q_{i} eB |}
\label{Ez}
\end{equation}
where $l$ = 0, 1, 2 ..... is the order of the Landau energy levels, $Q_{i}e$ is the fractional charge of quarks, and $p_z$ is the component of momentum along the z direction.\\

Furthermore, in the presence of the magnetic field, the three-dimensional integral is converted into the one-dimensional integral;
\begin{equation}
\int \frac{d^3 p}{(2\pi)^3} = \frac{|Q_i eB|}{2\pi} \int \frac{dp_z}{2\pi} 
\label{3to1}
\end{equation}

Since $q_T$ and $k_T$ are the momentum transfer of the mediated gluons, the integral over $q_T$ and $k_T$ present in Eq.~\ref{cs2} remains unaffected in the presence of magnetic field at the lowest order. \\

So, the fermionic contribution of Eq.~\ref{cs2} can be written as

\begin{equation}
\int \delta^4(p_3 + p_4 - p_1 -p_2) \delta^2(p_{3T} - p_{4T}) \frac{d^3p_3}{2 E_3} \frac{d^3p_4}{ 2 E_4} 
\label{delta1}
\end{equation}
The four-dimensional delta function function can be written as;
\begin{equation}
\delta^4(p_3 + p_4 - p_1 -p_2)  = \delta(E_3 + E_4 - E_1 -E_2) \delta^3(\vec{p}_3 + \vec{p}_4 - \vec{p}_1 -\vec{p}_2)
\label{delta2}
\end{equation}
In the presence of magnetic field the above Eq.~\ref{delta1} reduces to

\begin{align}
(|Q_{i}eB|)^2 (2\pi)^2 \sum_{l} \int \delta(E_{3z} + E_{4z} - E_{1z} -E_{2z}) \nonumber \\\delta(p_{3z} + p_{4z} - p_{1z} -p_{2z})  \delta(p_{3z} - p_{4z}) \frac{dp_{3z}}{2 E_{3z}} \frac{dp_{4z}}{ 2E_{4z}} 
\label{delta3}
\end{align}

Now Let's solve $\delta(E_{3z} + E_{4z} - E_{1z} -E_{2z})$
\begin{widetext}
\begin{align}
\delta(E_{3z} + E_{4z} - E_{1z} -E_{2z})  = \delta\bigg(\sqrt{p_{3z}^{2} + m_{3}^2 + 2l| Q_{3} eB |} + \sqrt{p_{4z}^{2} + m_{4}^2 + 2l| Q_{4} eB |} \nonumber \\ - \sqrt{p_{1z}^{2} + m_{1}^2 + 2l| Q_{1} eB |} -\sqrt{p_{2z}^{2} + m_{2}^2 + 2l| Q_{2} eB |}\bigg)
\label{delta4}
\end{align}
\end{widetext}

Using the property of the delta function, we have

\begin{equation}
\delta(f(x)) = \sum_{x_0} \frac{\delta(x - x_0)}{\bigg| \frac{df(x)}{dx} \bigg|_{x = x_0}}
\label{delta5}
\end{equation}
where $x_0$ is the root of the function f(x). \\
Let's denote the function inside the delta function as $f(p_{3z})$ 
\begin{align}
f(p_{3z}) = \sqrt{p_{3z}^{2} + m_{3}^2 + 2l| Q_{3} eB |} + \sqrt{p_{4z}^{2} + m_{4}^2 + 2l| Q_{4} eB |} \nonumber \\ - E_{1z} -E_{2z}
\label{fp3z}
\end{align}

Putting $f(p_{3z})$ = 0, we get two roots of  $p_{3z}$. Let's suppose $p^*_{3z}$ is the positive solution of $f(p_{3z})$; \\
and 
\begin{equation}
\frac{df(p_{3z})}{dp_{3z}}\vline_{p_{3z} = p^*_{3z}} = \bigg(\frac{E_{3z} + E_{4z}}{E_{3z} E_{4z}}\bigg)p^*_{3z}
\label{fp3zroot}
\end{equation}

Using the energy conservation $E_{3z} + E_{4z} = E_{1z} +E_{2z}$ and replacing Eq.~\ref{fp3zroot} in Eq.~\ref{delta3}, we obtain; 
\begin{align}
\frac{(2 \pi |Q_{i}eB|)^2}{4} \sum_{l}   \int \delta(p_{3z} - p^{*}_{3z}) \delta(p_{3z} + p_{4z} - p_{1z} -p_{2z}) \nonumber \\ \delta(p_{3z} - p_{4z}) \frac{ dp_{3z} dp_{4z} }{(E_{1z} +E_{2z})p^*_{3z}
}
\label{fp3zroot}
\end{align}

\begin{equation}
=  \frac{(2 \pi |Q_{i}eB|)^2}{4} \sum_{l}   \frac{1}{(E_{1z} +E_{2z})p_{3z}} \\
\label{fp3zroot}
\end{equation}
Using the property of the delta function, we obtain 
\begin{equation}
\int  \delta(p_{3z}+p_{4z}-p_{1z}-p_{2z})  \delta(p_{3z} - p_{4z})  dp_{3z} dp_{4z}  = 1
\label{fp3zroot}
\end{equation}
So, Eq.(11) becomes,
\begin{align}
\frac{d^2\sigma_{\lambda_3}}{d^2 x_T} = \sum_{l}   \sum_{\lambda_1, \lambda_2, \lambda_4} \frac{ (2 \pi |Q_{i}eB|)^2 c_{qq}}{16F (E_{1z} +E_{2z})p_{3z}} \int \frac{d^2 q_{T}}{(2\pi)^2} \frac{d^2 k_{T}}{(2\pi)^2}  \nonumber \\ e^{i\bf{(k_T - q_T) \cdot x_T}} \mathcal{M}_{f i}(q) \mathcal{M}_{f i}^{*}(k)
\label{cs3}
\end{align}
where  $\lambda_3$ = $\uparrow$ or $\downarrow$ denotes that spin of quark $q_1$ after the scattering is in the positive or negative direction of the normal ($\hat{n}$) to the reaction plane.\\

The Eq.~\ref{cs3} is the final expression for differential scattering cross section in the presence of the magnetic field.\\
The global quark polarization after the scattering is obtained using the relation given in Ref.~\cite{Liang:2007ma, Gao:2007bc};
\begin{equation}
P_q = \frac{<\Delta \sigma>}{<\sigma>}
\label{Pq}
\end{equation}
where,
\begin{equation}
<\Delta \sigma> = \int d^2x_T \; \; f_{qq} (x_T, b, Y, \sqrt{s}) \frac{d^2 \Delta \sigma}{d^2 x_T}
\label{deltasigma}
\end{equation}
\begin{equation}
< \sigma> = \int d^2x_T \; \; f_{qq} (x_T, b, Y, \sqrt{s}) \frac{d^2 \sigma}{d^2 x_T}
\label{sigma}
\end{equation}
where, $f_{qq} (x_T, b, Y, \sqrt{s})$ is the impact parameter ($x_T$) distribution. In this study, a simplistic uniform distribution was considered, such that;

\begin{equation}
f_{qq} (x_T, b, Y, \sqrt{s})  \propto  \theta (x) = 
\begin{cases}
1 , \; \; \; x >0 \\
0 , \; \; \; x <0 
\end{cases}
\label{fun}
\end{equation}

and $\frac{d^2 \Delta \sigma}{d^2 x_T}$, $\frac{d^2 \sigma}{d^2 x_T}$ denotes the polarized and unpolarized cross section at the fixed impact parameter. These are calculated using the mathematical expression;
\begin{equation}
\frac{d^2 \Delta \sigma}{d^2 x_T} = \frac{d^2 \sigma_{\uparrow}}{d^2 x_T} - \frac{d^2 \sigma_{\downarrow}}{d^2 x_T} 
\label{ddelsigma1}
\end{equation}
\begin{equation}
\frac{d^2 \sigma}{d^2 x_T} = \frac{d^2 \sigma_{\uparrow}}{d^2 x_T} + \frac{d^2 \sigma_{\downarrow}}{d^2 x_T}
\label{dsigma1}
\end{equation}
Here, the quark scattering is obtained through a static potential model, in which the quark incident in the z-direction is scattered by an effective static potential induced by another constituent of the QGP. Hence, in this case, we have

\begin{equation}
\mathcal{M}_{fi}(q) = g \; \bar{u}_{\lambda}(p+q) \cancel{A} (q) u(p)
\label{MMstar1}
\end{equation}

where $A(q) = (A_0 (q),0)$ and $A_0 (q) = g/(q^2 + m_D^{2})$ is the screened static potential with Debye screen mass $m_{D}$ \\
One can easily obtain, 
\begin{equation}
\mathcal{M}_{fi}(q) \mathcal{M}_{f i}^{*}(k) = g^2 A_0 (q) A_0 (k) \bar{u}_{\lambda}(p+q) (\cancel{p}+m_q) u_{\lambda}(p+k)
\label{MMstar2}
\end{equation}

where $p \simeq (E, -\bf{p})$. 
We choose $\hat{n}$ as the quantization axis of spin and denote the eigenvalue by $\lambda = \pm 1$. For small angle scattering, $q_T, k_T \simeq  m_D << E$, we obtain

\begin{align}
\mathcal{M}_{fi}(q) \mathcal{M}_{f i}^{*}(k) = \sum_{l} 4E_{z}^2 A_0 (q) A_0 (k) \nonumber \\\bigg[ 1 - i \lambda \frac{(\bf{q_T - k_T}) \cdot (\hat{n} \times \bf{p} )}{2 E_{z}(E_{z}+m_q)} \bigg]
\label{MMstar3}
\end{align}

Using Eq.~\ref{MMstar3} in Eq.~\ref{cs3}, the unpolarized and polarized scattering cross section in the presence of magnetic field can be obtained from the Eqs.~\ref{ddelsigma1} and ~\ref{dsigma1} and are given by,

\begin{equation}
\frac{d^2\sigma}{d^2 x_T} =  (\pi |Q_{i}eB|)^2 g^4 c_T \int \frac{d^2 q_{T}}{(2\pi)^2} \frac{d^2 k_{T}}{(2\pi)^2}  \frac{e^{i\bf{(k_T - q_T) \cdot x_T}}}{(q_T^2 + m_D^{2})(k_T^2 + m_D^{2})}
\label{ddelsigma2}
\end{equation}

\begin{align}
\frac{d^2 \Delta \sigma}{d^2 x_T} = i(\pi |Q_{i}eB|)^2 g^4 c_T \sum_{l} \int \frac{d^2 q_{T}}{(2\pi)^2} \frac{d^2 k_{T}}{(2\pi)^2}  \nonumber \\ \frac{ \bf{(k_T - q_T)} \cdot (\hat{n} \times \bf{p} ) e^{i\bf{(k_T - q_T) \cdot x_T}}}{2 E_{z}(E_{z}+m_q) (q_T^2 + m_D^{2})(k_T^2 + m_D^{2})}
\label{ddelsigma0}
\end{align}

Let's evaluate the integral
\begin{equation}
\int \frac{d^2 q_{T}}{(2\pi)^2} \frac{e^{i \bf{q_T \cdot x_T}}}{q_T^2 + m_D^{2}} = \int \frac{ q_{T} dq_{T}}{2\pi} \frac{J_0 (q_T x_T)}{q_T^2 + m_D^{2}} = \frac{K_0 (m_D x_T)}{2 \pi}
\label{int1}
\end{equation}
Similarly, 
\begin{equation}
\int \frac{d^2 k_{T}}{(2\pi)^2} \frac{e^{i \bf{k_T \cdot x_T}}}{k_T^2 + m_D^{2}} = \frac{K_0 (m_D x_T)}{2 \pi}
\label{int2}
\end{equation}
Using these integral expression in Eq.~\ref{ddelsigma2} and Eq.~\ref{ddelsigma0}, we have
\begin{equation}
\frac{d^2\sigma}{d^2 x_T} = \frac{( |Q_{i}eB|)^2g^4 c_T}{4} K_0^2 (m_D x_T)
\label{dsigma3}
\end{equation}

\begin{align}
\frac{d^2 \Delta \sigma}{d^2 x_T} = \sum_{l} \frac{(|Q_{i}eB|)^2g^4 c_T}{4}  \frac{(\bf{p}  \times \hat{n}) \cdot x_T}{2E_{z}(E_{z}+m_q)} \nonumber \\ m_D K_0 (m_D x_T) K_1 (m_D x_T)
\label{ddelsigma3}
\end{align}
where $J_0$ and $K_0$ are the Bessel and modified Bessel functions, respectively.

Substituting Eq.~\ref{dsigma3} and Eq.~\ref{ddelsigma3} in Eq.~\ref{Pq} and averaging over the relative angle between the parton $x_T$ and nuclear impact parameter b from -$\pi/2$ to +$\pi/2$ and over $x_T$, one can obtain the global quark polarization,

\begin{equation}
P_q = \sum_{l} \frac{-\pi m_D \bf{p}}{2E_{z}(E_{z}+m_q)}
\label{Pq2}
\end{equation}
The Eq.~\ref{Pq2} is the final expression of global quark polarization in the presence of magnetic field.

\vspace{10.005em}

\begin{thebibliography}{}

\bibitem{ALICE:2016fzo}
J.~Adam \textit{et al.} [ALICE Collaboration],
Nature Phys. \textbf{13}, 535 (2017).

\bibitem{CMS:2016fnw}
V.~Khachatryan \textit{et al.} [CMS Collaboration],
Phys. Lett. B \textbf{765}, 193 (2017).

\bibitem{OrtizVelasquez:2013ofg}
A.~Ortiz Velasquez, P.~Christiansen, E.~Cuautle Flores, I.~Maldonado Cervantes and G.~Pai\'c,
Phys. Rev. Lett. \textbf{111}, 042001 (2013).

\bibitem{Zhang:2021phk}
C.~Zhang, A.~Behera, S.~Bhatta and J.~Jia,
Phys. Lett. B \textbf{822}, 136702 (2021).

\bibitem{Sahoo:2023oid}
B.~Sahoo, D.~Sahu, S.~Deb, C.~R.~Singh and R.~Sahoo,
Phys. Rev. C \textbf{109}, 034910 (2024).


\bibitem{Liang:2004ph}
Z.~T.~Liang and X.~N.~Wang,
Phys. Rev. Lett. \textbf{94}, 102301 (2005).
[erratum: Phys. Rev. Lett. \textbf{96} (2006), 039901]

\bibitem{Liang:2004xn}
Z.~T.~Liang and X.~N.~Wang,
Phys. Lett. B \textbf{629}, 20 (2005).

\bibitem{Liang:2007ma}
Z.~t.~Liang,
J. Phys. G \textbf{34}, S323-330 (2007).

\bibitem{Gao:2007bc}
J.~H.~Gao, S.~W.~Chen, W.~t.~Deng, Z.~T.~Liang, Q.~Wang and X.~N.~Wang,
Phys. Rev. C \textbf{77}, 044902 (2008).
\bibitem{Gao:2012ix}
J.~H.~Gao, Z.~T.~Liang, S.~Pu, Q.~Wang and X.~N.~Wang,
Phys. Rev. Lett. \textbf{109}, 232301 (2012).

\bibitem{Teryaev:2017wlm}
O.~V.~Teryaev and V.~I.~Zakharov,
Phys. Rev. D \textbf{96}, 096023 (2017).

\bibitem{Sheng:2019kmk}
X.~L.~Sheng, L.~Oliva and Q.~Wang,
Phys. Rev. D \textbf{101}, 096005 (2020).
[erratum: Phys. Rev. D \textbf{105}, 099903 (2022).]

\bibitem{Sheng:2022wsy}
X.~L.~Sheng, L.~Oliva, Z.~T.~Liang, Q.~Wang and X.~N.~Wang,
Phys. Rev. Lett. \textbf{131}, 042304 (2023).

\bibitem{Becattini:2017gcx}
F.~Becattini and I.~Karpenko,
Phys. Rev. Lett. \textbf{120}, 012302 (2018).

\bibitem{STAR:2019erd}
J.~Adam \textit{et al.} [STAR Collaboration],
Phys. Rev. Lett. \textbf{123}, 132301 (2019).

\bibitem{FNAL-E704:1991ovg}
D.~L.~Adams \textit{et al.} [FNAL-E704],
Phys. Lett. B \textbf{264}, 462(1991).

\bibitem{Ipp:2007ng}
A.~Ipp, A.~Di Piazza, J.~Evers and C.~H.~Keitel,
Phys. Lett. B \textbf{666}, 315 (2008).

\bibitem{Becattini:2013fla}
F.~Becattini, V.~Chandra, L.~Del Zanna and E.~Grossi,
Annals Phys. \textbf{338}, 32 (2013). 

\bibitem{Fang:2016vpj}
R.~h.~Fang, L.~g.~Pang, Q.~Wang and X.~n.~Wang,
Phys. Rev. C \textbf{94}, 024904 (2016).

\bibitem{Xie:2017upb}
Y.~Xie, D.~Wang and L.~P.~Csernai,
Phys. Rev. C \textbf{95}, 031901 (2017).

\bibitem{STAR:2007ccu}
B.~I.~Abelev \textit{et al.} [STAR Collaboration],
Phys. Rev. C \textbf{76}, 024915 (2007).

\bibitem{STAR:2017ckg}
L.~Adamczyk \textit{et al.} [STAR Collaboration],
Nature \textbf{548}, 62 (2017).

\bibitem{STAR:2018gyt}
J.~Adam \textit{et al.} [STAR Collaboration],
Phys. Rev. C \textbf{98}, 014910 (2018).

\bibitem{STAR:2021beb}
M.~S.~Abdallah \textit{et al.} [STAR Collaboration],
Phys. Rev. C \textbf{104}, L061901 (2021).


\bibitem{STAR:2023nvo}
M.~I.~Abdulhamid \textit{et al.} [STAR Collaboration],
Phys. Rev. C \textbf{108}, 014910 (2023).

\bibitem{Karpenko:2016jyx}
I.~Karpenko and F.~Becattini,
Eur. Phys. J. C \textbf{77}, 213 (2017).

\bibitem{Alzhrani:2022dpi}
S.~Alzhrani, S.~Ryu and C.~Shen,
Phys. Rev. C \textbf{106}, 014905 (2022).

\bibitem{Vitiuk:2019rfv}
O.~Vitiuk, L.~V.~Bravina and E.~E.~Zabrodin,
Phys. Lett. B \textbf{803}, 135298 (2020).

\bibitem{Jiang:2016woz}
Y.~Jiang, Z.~W.~Lin and J.~Liao,
Phys. Rev. C \textbf{94}, 044910  (2016).
[erratum: Phys. Rev. C \textbf{95}, 049904 (2017).]

\bibitem{Sun:2017xhx}
Y.~Sun and C.~M.~Ko,
Phys. Rev. C \textbf{96}, 024906 (2017).

\bibitem{Ivanov:2019ern}
Y.~B.~Ivanov, V.~D.~Toneev and A.~A.~Soldatov,
Phys. Rev. C \textbf{100}, 014908 (2019).

\bibitem{Ivanov:2022ble}
Y.~B.~Ivanov and A.~A.~Soldatov,
Phys. Rev. C \textbf{105}, 034915  (2022).

\bibitem{Pradhan:2023rvf}
K.~K.~Pradhan, B.~Sahoo, D.~Sahu and R.~Sahoo,
Eur. Phys. J. C \textbf{84}, 936 (2024).

\bibitem{Wei:2018zfb}
D.~X.~Wei, W.~T.~Deng and X.~G.~Huang,
Phys. Rev. C \textbf{99}, 014905 (2019).

\bibitem{Becattini:2024uha}
F.~Becattini, M.~Buzzegoli, T.~Niida, S.~Pu, A.~H.~Tang and Q.~Wang,
Int. J. Mod. Phys. E \textbf{33}, 2430006  (2024).

\bibitem{Guo:2021udq}
Y.~Guo, J.~Liao, E.~Wang, H.~Xing and H.~Zhang,
Phys. Rev. C \textbf{104}, L041902 (2021).

\bibitem{Csernai:2013bqa}
L.~P.~Csernai, V.~K.~Magas and D.~J.~Wang,
Phys. Rev. C \textbf{87}, 034906 (2013).

\bibitem{Csernai:2014ywa}
L.~P.~Csernai, D.~J.~Wang, M.~Bleicher and H.~St\"ocker,
Phys. Rev. C \textbf{90}, 021904 (2014).

\bibitem{Pang:2016igs}
L.~G.~Pang, H.~Petersen, Q.~Wang and X.~N.~Wang,
Phys. Rev. Lett. \textbf{117}, 192301 (2016).

\bibitem{Li:2017slc}
H.~Li, L.~G.~Pang, Q.~Wang and X.~L.~Xia,
Phys. Rev. C \textbf{96}, 054908 (2017).  

\bibitem{Fu:2020oxj}
B.~Fu, K.~Xu, X.~G.~Huang and H.~Song,
Phys. Rev. C \textbf{103}, 024903 (2021).

\bibitem{Karpenko:2017lyj}
I.~Karpenko and F.~Becattini,
Nucl. Phys. A \textbf{967}, 764 (2017).

\bibitem{Becattini:2020ngo}
F.~Becattini and M.~A.~Lisa,
Ann. Rev. Nucl. Part. Sci. \textbf{70}, 395 (2020).

\bibitem{Xie:2020}
Xie, Y., Wang, D. and Csernai, L.P., 
Eur. Phys. J. C \textbf{80}, 39 (2020). 


\bibitem{ALICE:2019onw}
S.~Acharya \textit{et al.} [ALICE Collaboration],
Phys. Rev. C \textbf{101}, 044611 (2020). [erratum: Phys. Rev. C \textbf{105}, 029902 (2022).]

\bibitem{STAR:2020xbm}
J.~Adam \textit{et al.} [STAR Collaboration],
Phys. Rev. Lett. \textbf{126}, 162301 (2021).

\bibitem{Alpatov:2020iev}
E.~Alpatov [STAR Collaboration],
J. Phys. Conf. Ser. \textbf{1690}, 012120 (2020).

\bibitem{ALICE:2020iev}
S.~Acharya \textit{et al.}, [ALICE Collaboration],
Phys. Lett. B \textbf{815}, 136146 (2021).
\bibitem{ALICE:2023jad}
S.~Acharya \textit{et al.}, [ALICE Collaboration],
Phys. Rev. Lett. \textbf{131}, 042303 (2023).

\bibitem{STAR:2013iae}
L.~Adamczyk \textit{et al.}, [STAR Collaboration],
Phys. Lett. B \textbf{739}, 180 (2014).

\bibitem{STAR:2020igu}
J.~Adam \textit{et al.}, [STAR Collaboration],
Phys. Rev. D \textbf{102}, 092009 (2020).


\bibitem{STAR:2022fan}
M.~S.~Abdallah \textit{et al.} [STAR Collaboration],
Nature \textbf{614} 244 (2023). 

\bibitem{Singha:2020qns}
S.~Singha [STAR Collaboration],
Nucl. Phys. A \textbf{1005}, 121733 (2021).

\bibitem{STAR:2008lcm}
B.~I.~Abelev \textit{et al.} [STAR Collaboration],
Phys. Rev. C \textbf{77}, 061902 (2008).

\bibitem{ALICE:2019aid}
S.~Acharya \textit{et al.} [ALICE Collaboration],
Phys. Rev. Lett. \textbf{125}, 012301 (2020).

\bibitem{Singh:2018uad}
R.~Singh [ALICE Collaboration],
Nucl. Phys. A \textbf{982}, 515 (2019).

\bibitem{Mohanty:2021vbt}
B.~Mohanty, S.~Kundu, S.~Singha and R.~Singh,
Mod. Phys. Lett. A \textbf{36}, 2130026  (2021).

\bibitem{Csernai:2018yok}
L.~P.~Csernai, J.~I.~Kapusta and T.~Welle,
Phys. Rev. C \textbf{99}, 021901 (2019).

\bibitem{Becattini:2021iol}
F.~Becattini, M.~Buzzegoli, G.~Inghirami, I.~Karpenko and A.~Palermo,
Phys. Rev. Lett. \textbf{127}, 272302 (2021).


\bibitem{Wu:2019eyi}
H.~Z.~Wu, L.~G.~Pang, X.~G.~Huang and Q.~Wang,
Phys. Rev. Research. \textbf{1}, 033058 (2019).

\bibitem{Xia:2020tyd}
X.~L.~Xia, H.~Li, X.~G.~Huang and H.~Zhong Huang,
Phys. Lett. B \textbf{817}, 136325  (2021).

\bibitem{Skokov:2009qp}
V.~Skokov, A.~Y.~Illarionov and V.~Toneev,
Int. J. Mod. Phys. A \textbf{24}, 5925 (2009).

\bibitem{Sahoo:2023xnu}
B.~Sahoo, C.~R.~Singh, D.~Sahu, R.~Sahoo and J.~e.~Alam,
Eur. Phys. J. C \textbf{83}, 873 (2023).

\bibitem{Sahoo:2023vkw}
B.~Sahoo, K.~K.~Pradhan, D.~Sahu and R.~Sahoo,
Phys. Rev. D \textbf{108}, 074028 (2023).


\bibitem{Yang:2017sdk}
Y.~G.~Yang, R.~H.~Fang, Q.~Wang and X.~N.~Wang,
Phys. Rev. C \textbf{97}, 034917 (2018).

\bibitem{Sheng:2020ghv}
X.~L.~Sheng, Q.~Wang and X.~N.~Wang,
Phys. Rev. D \textbf{102}, 056013 (2020).

\bibitem{Becattini:2016gvu}
F.~Becattini, I.~Karpenko, M.~Lisa, I.~Upsal and S.~Voloshin,
Phys. Rev. C \textbf{95}, 054902 (2017).

\bibitem{Han:2017hdi}
Z.~Z.~Han and J.~Xu,
Phys. Lett. B \textbf{786}, 255 (2018).

\bibitem{Guo:2019joy}
Y.~Guo, S.~Shi, S.~Feng and J.~Liao,
Phys. Lett. B \textbf{798}, 134929 (2019).

\bibitem{Xu:2022hql}
K.~Xu, F.~Lin, A.~Huang and M.~Huang,
Phys. Rev. D \textbf{106}, L071502 (2022).

\bibitem{Guo:2019mgh}
X.~Guo, J.~Liao and E.~Wang,
Sci. Rep. \textbf{10}, 2196 (2020).

\bibitem{Wu:2023dfz}
S.~Wu and Y.~Xie,
Eur. Phys. J. A \textbf{59}, 108 (2023).

\bibitem{Ayala:2020soy}
A.~Ayala, M.~A.~Ayala Torres, E.~Cuautle, I.~Dom\'\i{}nguez, M.~A.~Fontaine Sanchez, I.~Maldonado, E.~Moreno-Barbosa, P.~A.~Nieto-Mar\'\i{}n, M.~Rodr\'\i{}guez-Cahuantzi and J.~Salinas, \textit{et al.}
Phys. Lett. B \textbf{810}, 135818 (2020).


\bibitem{Dong:2021gnb}
L.~Dong, Y.~Guo, A.~Islam and M.~Strickland,
Phys. Rev. D \textbf{104}, 096017 (2021).

\bibitem{Dong:2022mbo}
L.~Dong, Y.~Guo, A.~Islam, A.~Rothkopf and M.~Strickland,
JHEP \textbf{09}, 200 (2022).

\bibitem{Strickland:2011aa}
M.~Strickland and D.~Bazow,
Nucl. Phys. A \textbf{879}, 25 (2012).

\bibitem{Romatschke:2003ms}
P.~Romatschke and M.~Strickland,
Phys. Rev. D \textbf{68}, 036004 (2003).
\bibitem{Romatschke:2004jh}
P.~Romatschke and M.~Strickland,
Phys. Rev. D \textbf{70}, 116006 (2004).
\bibitem{Strickland:lectures}
M. Strickland, Anisotropic hydrodynamics: Three lectures, Acta Phys. Pol. B \textbf{45}, 2355 (2014).

\bibitem{Karmakar:2018aig}
B.~Karmakar, A.~Bandyopadhyay, N.~Haque and M.~G.~Mustafa,
Eur. Phys. J. C \textbf{79}, 658 (2019).

\bibitem{Zhang:2020efz}
H.~X.~Zhang, J.~W.~Kang and B.~W.~Zhang,
Eur. Phys. J. C \textbf{81}, 623 (2021).

\bibitem{Bandyopadhyay:2017cle}
A.~Bandyopadhyay, B.~Karmakar, N.~Haque and M.~G.~Mustafa,
Phys. Rev. D \textbf{100}, 034031 (2019).

\bibitem{Singh:2017nfa}
B.~Singh, L.~Thakur and H.~Mishra,
Phys. Rev. D \textbf{97}, 096011 (2018).

\bibitem{Haque:2014rua}
N.~Haque, A.~Bandyopadhyay, J.~O.~Andersen, M.~G.~Mustafa, M.~Strickland and N.~Su,
JHEP \textbf{05}, 027 (2014).

\bibitem{Zhao:2022cvl}
J.~Zhao and B.~Chen,
Phys. Rev. C \textbf{107}, 044909 (2023).

\bibitem{Lafferty:2019jpr}
D.~Lafferty and A.~Rothkopf,
Phys. Rev. D \textbf{101}, 056010  (2020).

\bibitem{Huang:2021ysc}
G.~Huang and P.~Zhuang,
Phys. Rev. D \textbf{104}, 074001  (2021).

\bibitem{Bellac}
M. Le Bellac, Thermal Field Theory (Cambridge Univer-
sity Press, Cambridge, England, 1996).


\bibitem{Khan:2021syq}
S.~A.~Khan, M.~Hasan and B.~K.~Patra,
Nucl. Phys. A \textbf{1034}, 122643 (2023).

\bibitem{Bonati:2017uvz}
C.~Bonati, M.~D'Elia, M.~Mariti, M.~Mesiti, F.~Negro, A.~Rucci and F.~Sanfilippo,
Phys. Rev. D \textbf{95}, 074515 (2017).

\bibitem{Bandyopadhyay:2016fyd}
A.~Bandyopadhyay, C.~A.~Islam and M.~G.~Mustafa,
Phys. Rev. D \textbf{94}, 114034 (2016).


\bibitem{Bandyopadhyay:2019pml}
A.~Bandyopadhyay, R.~L.~S.~Farias, B.~S.~Lopes and R.~O.~Ramos,
Phys. Rev. D \textbf{100}, 076021 (2019).

\bibitem{Zhang:2018vyr}
J.~w.~Zhang, H.~h.~Li, F.~l.~Shao and J.~Song,
Chin. Phys. C \textbf{44}, 014101 (2020).

\bibitem{Ryu:2021lnx}
S.~Ryu, V.~Jupic and C.~Shen,
Phys. Rev. C \textbf{104}, 054908 (2021).

\bibitem{Wu:2022mkr}
X.~Y.~Wu, C.~Yi, G.~Y.~Qin and S.~Pu,
Phys. Rev. C \textbf{105}, 064909 (2022).

\bibitem{Kolomeitsev:2018svb}
E.~E.~Kolomeitsev, V.~D.~Toneev and V.~Voronyuk,
Phys. Rev. C \textbf{97}, 064902 (2018).

\bibitem{Xia:2019fjf}
X.~L.~Xia, H.~Li, X.~G.~Huang and H.~Z.~Huang,
Phys. Rev. C \textbf{100}, 014913 (2019).

\bibitem{Voronyuk:2023vyu}
V.~Voronyuk, N.~S.~Tsegelnik and E.~E.~Kolomeitsev,
Phys. Rev. C \textbf{111}, 034907 (2025).

\bibitem{Yi:2021unq}
C.~Yi, S.~Pu, J.~H.~Gao and D.~L.~Yang,
Phys. Rev. C \textbf{105}, 044911  (2022).

\end{thebibliography}
\end{document}